\documentclass[aps,showpacs,floatfix,nofootinbib,reprint,twocolumn]{revtex4}
\usepackage{graphicx}
\usepackage{epsf}
\usepackage{wrapfig}
\usepackage{epsfig}
\usepackage{sublabel}
\usepackage{epsfig}
\usepackage{verbatim}
\usepackage{amsmath}
\usepackage{color}
\usepackage{soul}
\usepackage{ulem}
\usepackage{multirow}

\def\b0{{\mbox{\boldmath$0$}}}

\usepackage{graphicx}
\usepackage{epsf}
\usepackage{wrapfig}
\usepackage{epsfig}
\def\Vec#1{\mbox{\boldmath $#1$}}

\def\beq{\begin{equation}}
\def\eeq{\end{equation}}

\def\beqy{\begin{eqnarray}}
\def\eeqy{\end{eqnarray}}
\def\ie{\textit{i.e. }}

\def \b #1{ {\bf #1}}
\newcommand{\be}{\begin{eqnarray}}
\newcommand{\ee}{\end{eqnarray}}

\def \b #1{ {\bf #1}}

\def \b #1{ {\bf #1}}
     \font\tenbifull=cmmib10 scaled 1200 
     \font\tenbimed=cmmib9
     \font\tenbismall=cmmib7
\mathchardef\bbkappa="7114 \mathchardef\bbrho="711A
\mathchardef\bbsigma="711B \mathchardef\bbtau="711C
\mathchardef\bbvarrho="7125 \mathchardef\bbvarsigma="7126
\mathchardef\bbxi="7118

\begin{document}
\vskip 2mm
\date{\today}\vskip 2mm
\title{Universality of nucleon-nucleon short-range correlations: the factorization property
  of the nuclear wave function, the relative and center-of-mass momentum distributions,
  and the nuclear contacts}
\author{M. Alvioli$^1$\footnote{Supported by grants provided by the Regione Umbria, under contract
    POR-FESR Umbria 20072013, asse ii, attivit\`a a1, azione 5, and by the Dipartimento
    della Protezione Civile, Italy.}}
\email{massimiliano.alvioli@irpi.cnr.it}
\author{C. Ciofi degli Atti$^{2}$}
\email{ciofi@pg.infn.it}
\author{\mbox{H. Morita$^3$}}
\email{hiko@webmail.sgu.ac.jp} \affiliation{$^1$CNR-IRPI, Istituto
  di Ricerca per la Protezione Idrogeologica,
  Via Madonna Alta 126, I-06128, Perugia, Italy\\
  $^2$Istituto Nazionale di Fisica Nucleare, Sezione di Perugia,\\
  Department of Physics, University of Perugia,
  Via A. Pascoli, I-06123, Italy\\
  $^3$\mbox{Sapporo Gakuin University, Bunkyo-dai 11, Ebetsu 069-8555, Hokkaido, Japan}}
\vskip 2mm
\begin{abstract}
  {{\bf Background}:
    the  two-nucleon momentum distributions of  nucleons $N_1$ and $N_2$ in a nucleus $A$,
    $n_A^{N_1N_2}({\bf k}_{rel},{\bf K}_{c.m.})$, is a relevant quantity that determines
    the probability  to find the two nucleons
    with relative momentum ${\bf k}_{rel}$ and center-of-mass (c.m.) momentum
    ${\bf K}_{c.m.}$; at large values of the relative momentum and, at the same time,
    small values of the c.m. momentum,  $n_A^{N_1N_2}({\bf k}_{rel},{\bf K}_{c.m.})$ provides information on the short-range
    structure of nuclei.\\
    \indent {\bf Purpose}: calculation of  the momentum distributions of proton-neutron and proton-proton
    pairs in $^3$He, $^4$He, $^{12}$C, $^{16}$O and $^{40}$Ca, in correspondence of various values of
    ${\bf k}_{rel}$ and ${\bf K}_{c.m.}$.\\
    \indent {\bf Methods}:  the momentum distributions for $A>4$ nuclei are calculated as a function of the relative, $k_{\text{rel}}$,
    and center of mass, $K_{\text{c.m.}}$, momenta and relative  angle $\Theta$, within a  linked cluster many-body
    expansion approach, based upon realistic local two-nucleon interaction of the Argonne
    family and variational wave functions featuring central, tensor and spin-isospin correlations.\\
    \indent {\bf Results}: independently of the mass number $A$, at values of the relative momentum
    $k_{\text{rel}}\gtrsim 1.5 \sim 2 fm^{-1}$ the momentum distributions exhibit the property of factorization,
    $n_A^{N_1N_2}({\bf k}_{rel},{\bf K}_{c.m.}) \simeq n_{rel}^{N_1N_2}({k}_{rel})n_{c.m.}^{N_1N_2}({K}_{c.m.})$,
    in particular for $pn$ back-to-back (BB) pairs one has $n_A^{pn}({k}_{rel},{K}_{c.m.}=0)\simeq C_A^{pn}n_D(k_{rel})n_{c.m.}^{pn}(K_{c.m.}=0)$
    where $n_D$ is
    the deuteron momentum distribution,  $n_{c.m.}^{pn}(K_{c.m.}=0)$ the c.m. motion momentum distribution of the pair and
    $C_A^{pn}$ the $pn$ {\it nuclear contact} measuring the number
    of BB $pn$ pairs  with deuteron-like momenta (${\bf k}_p \simeq -{\bf k}_n, {\bf K}_{c.m.}=0$),}. The values
    of the $pn$ {\it nuclear contact} are extracted
    from the general properties of the two-nucleon momentum distributions corresponding to $K_{c.m.}=0$. The $K_{c.m.}$-integrated $pn$
    momentum distributions exhibit  the property $n_A^{pn}(k_{rel})\simeq C_A^{pn}n_D(k_{rel})$ but only at very high
    value of $k_{rel}\gtrsim 3.5 \sim 4\,fm^{-1}$. The theoretical ratio of the $pp/pn$ momentum distributions of $^4$He and $^{12}$C
    and the calculated c.m. motion momentum distributions are in  agreement
    with recent  experimental data.
\end{abstract}
\maketitle
\newpage
\section{Aim and introduction}\label{Sec:1}
The investigation of short-range correlations (SRCs) in nuclei is ultimately aimed at unveiling the details of in-medium short-range
nucleon-nucleon (NN) dynamics,  a relevant physics issue that cannot be answered by scattering experiments of two free nucleons (see recent
review papers on the subject \cite{Frankfurt:2008zv,Arrington:2011xs,Alvioli:2013qyz, CiofidegliAtti_book,Atti:2015eda}). A reliable way to
gather  information on SRCs would be to detect significant deviations of proper experimental data   (e.g. electro-disintegration processes
off nuclei) from  theoretical predictions based upon {\it ab initio} solutions of the nuclear many-body problem,  obtained from various  NN
interactions  differing in the short-range part. In practice such an approach faces several problems because it implies the exact
calculation of the  ground- and continuum-state wave functions of the target nucleus under investigation; concerning the former, relevant
progress has recently been made
  to obtain {\it ab initio}
solutions of the non relativistic Schr\"{o}dinger equation, but, unfortunately,  the treatment of the continuum spectrum of the target
nucleus
 is still model-dependent,  with the only exception of  those processes involving the
 two- and three-nucleon systems; for  complex nuclei  approximations are
unavoidable, with the simplest one being  the plane wave impulse approximation (PWIA) which leads, in the case of, e. g., a process
$A(e,e'N)X$,  to a factorized cross sections depending upon the  elementary electron-nucleon cross section and the one-nucleon  spectral
function $P_A(E,k)$  which describes  the momentum ($k \equiv |{\bf k}|$) and removal  energy ($E$) distributions of a nucleon in  nucleus
$A$ (in a process $A(e,e'N_1N_2)X$) the factorized cross section will depend upon the two-nucleon spectral function, etc.).  Even if the
PWIA requires  corrections due to the final state interaction (FSI) and possible effects from non-nucleonic degrees of freedom, the
detection of high momentum and high removal energy effects may represents evidence of ground-state SRCs. It is for this reason that  during
the last  few years, the calculation of the nuclear momentum  distributions and spectral function has attracted an increasing interest. The
one-nucleon, $n_A({k}_1)$,   and two-nucleon, $n_A({\bf k}_1,{\bf k}_2)$, momentum distributions  of few-nucleon systems  ($A \leq 4$) have
been obtained {\it ab initio}  \cite{Suzuki:2008cy,Feldmeier:2011qy,Roth:2010bm,Alvioli:2011aa,Schiavilla:2006xx} within different
theoretical approaches and using realistic NN interactions, whereas for  $A\leq 12$  exact  variational Monte Carlo (VMC) calculations have
recently  been  performed \cite{Wiringa:2013ala}. For nuclei  with $A > 12$, VMC calculations of the momentum distribution are not yet
feasible, therefore, also in light of future experimental developments, alternative approaches, even if of lower quality than VMC ones, but
still maintaining a realistic link to the underlying NN interactions, should be pursued. A serious candidate in this respect would be an
advanced linked cluster expansion approach with correlated wave functions, including a large class of Yvon-Mayer diagrams
\cite{Gaudin:1971zz,Bohigas,AriasdeSaavedra:2007byz},  for they have been shown to produce realistic results of one-nucleon momentum
distributions \cite{Alvioli:2005cz,Alvioli:2012qa,Alvioli:2007zz}  in reasonable agreement with the more advanced VMC  calculations. All of
these calculations, though being performed within different many-body approaches, produce similar results demonstrating a universal
(A-independent) character of  in-medium NN short-range dynamics, in that the mean-field (MF) approach breaks down when the relative
distance $r\equiv |{\Vec r}_1 - {\Vec r}_2|$ between two generic nucleons "1" and "2" is of the order of $r \lesssim 1.3 -1.5$ fm, with the
two-nucleon density distribution exhibiting the so called {\it correlation hole} which, apart from trivial normalization factors, turns out
to be independent on the mass $A$ of the nucleus and similar to the deuteron one. SRCs give rise to high momentum components that are
lacking in a mean-field approach and turned out to depend upon the relative orbital momentum (L) and  the total spin (S) and isospin (T) of
the $NN$ pair, as well as upon the value of the pair center-of-mass (c.m.)
momentum. SRCs give rise to peculiar configurations of the
nuclear wave function in momentum space, e.g. the ones when a high momentum nucleon is mostly balanced by another  nucleon with similar and
opposite value of the momentum (the {\it back-to-back} (BB) configuration) and not by the $A-1$ nucleon, as in the case of a mean-field
configuration \cite{Frankfurt:1988nt}. Thus, within a PWIA picture, if a correlated nucleon, with momentum ${\Vec k}_1$, acquiring a
momentum ${\Vec q}$ from an external probe, leaves the nucleus without any final-state interaction (FSI) and  is detected with  momentum
${\Vec p}={\Vec k}_1+{\Vec q}$, the partner nucleon should be emitted with momentum ${\Vec k}_2 \simeq {\Vec p}_{m} =-{\Vec k}_1$, where
the measurable  momentum  ${\Vec p}_{m}$ is the {\it missing momentum} ${\Vec p}_{m} ={\Vec q}-{\Vec p}$. Such a  basic picture of {\it
back-to-back}  short-range correlated (SRCd) nucleons has been  recently improved to a large extent  by taking into account the FSI of the
struck nucleon by advanced methods (see {\it e.g.} Refs. \cite{Frankfurt:1996xx,Mardor:1992sb} and \cite{Ryckebusch:2003fc,Cosyn:2007er})
and by considering the effects due to the center-of-mass motion of the pair,
which makes ${\Vec k}_2 \neq- {\Vec k}_1$, and the effects due to
the (ST) dependence. The  underlying dynamics of  SRCs has been theoretically explained by advanced many-body theories, \textit{e.g.}  by
the Brueckner-Bethe-Goldstone approach for nuclear matter \cite{Baldo:1900zz} and  by  exact  few-nucleon approaches in case of $^3$He and
$^4$He \cite{CiofidegliAtti:2010xv}, with both approaches   demonstrating  that  two-nucleon correlations arise from a general property of
the many-body wave
 function, namely its factorized form in those configurations where a pair of nucleons
 has, at the same time,  a  large
value of the  two-nucleon relative momentum $\Vec{k}_{\text{rel}}$ and a low value of
the c.m. momentum $\Vec{K}_{\text{c.m.}}$, in agreement with the phenomenological
assumption of Ref. \cite{CiofidegliAtti:1995qe}. The
presence of SRCs in nuclei and their  basic back-to-back  nature
 have eventually been  experimentally demonstrated \cite {Tang:2002ww,Piasetzky:2006ai,Shneor:2007tu,
 Subedi:2008zz,Korover:2014dma,Hen:2014nza}, but a detailed theoretical and
 experimental information  through the
periodic Table of Elements of  their  isospin, angular and c.m. momentum dependencies  remains to be obtained. To contribute to this challenge  in
the present paper  the results of calculations of the following quantities, pertaining to nuclei $^3$He, $^4$He, $^{12}$C, $^{16}$O and
$^{40}$Ca, will be presented: (i) the two-nucleon momentum distribution $n_A^{N_1N_2}$ of the proton-neutron ($pn$) and proton-proton
($pp$) pairs in correspondence of different values of the c.m.
 and the the relative  momenta of the pair and
the angle  $\Theta$ between them; (ii) the number of short-range correlated
 $pp$ and $pn$ pairs represented by the integral of the
various types of momentum distributions in a finite momentum range; (iii) the ratio of the $pn$ to $pp$ correlated pairs \textit{vs} the
relative momentum $k_{rel}$. Particular attention is devoted to the comparison of the two-nucleon momentum distributions of complex nuclei
with the deuteron momentum distribution,  in order to clarify whether and to which extent the short-range dynamics of a  free bound  $pn$
system will differ from the short-range dynamics of a $pn$ pair embedded in the medium. Calculations have been  performed with realistic
nuclear wave functions \cite{Alvioli:2005cz,Kievsky:1992um,Kievsky:2010zr,Akaishi:1987} obtained from the solution of the Schr\"odinger
equation with realistic $NN$ interactions, namely the AV18 \cite{Wiringa:1994wb} and $AV8'$ \cite{Pudliner:1997ck} interactions. Various
properties of the momentum distributions and various relations between them are illustrated, which further demonstrate the relevant
property of the nuclear wave function  in the correlation region, \ie its factorized form. The quantity (the {\it nuclear contact})
measuring the number of deuteron-like pairs in nuclei is extracted from the general properties of the $pn$ momentum distributions. The
structure of the paper is as follows: in Section \ref{Sec:2} the general definitions of the two-nucleon momentum distributions and their
SRCd parts are given; the calculation  of the momentum distributions and the universal, A-independent behavior of their SRCd parts, are
presented in Section \ref{Sec:3}; the general validity of the factorization property in the SRC region is proved in Section \ref{Sec:4};
the number of SRCd $pn$ and $pp$ pairs in various regions of $k_{rel}$ and $K_{c.m.}$ are given in Section \ref{Sec:5}; the comparison
between the available experimental data with theoretical predictions is presented in Section \ref{Sec:6}; the Summary and Conclusions are
illustrated in Section \ref{Sec:7}.
\section{General definitions} \label{Sec:2}
\begin{figure*}[!htp] 
\vskip 1.0cm
  \centerline{\includegraphics[width=1.0\textwidth]{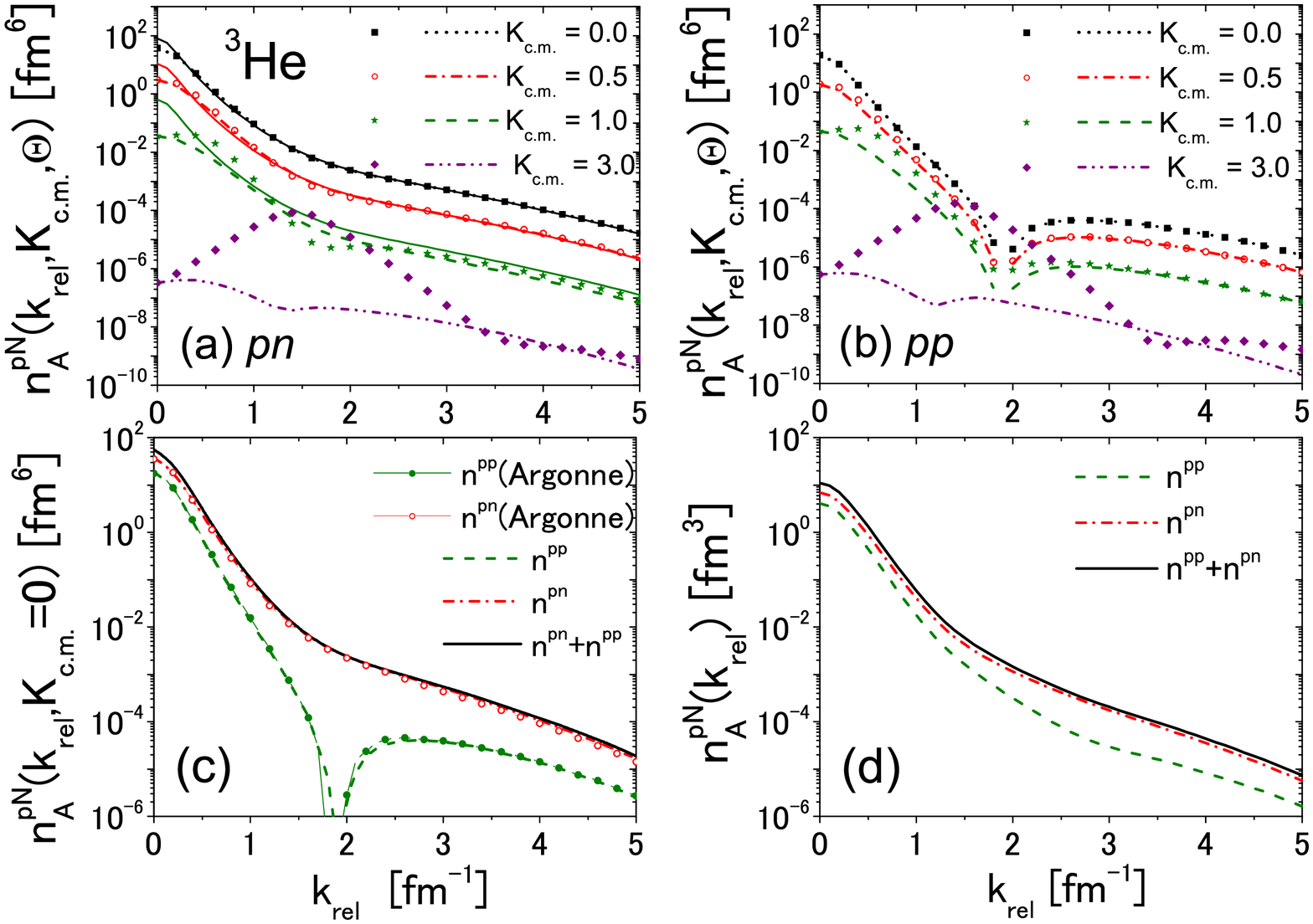}}
  \caption{(Color online): (a): the  two-nucleon momentum distributions of $pn$
      pairs in  $^3$He  \textit{vs.} the relative momentum $k_{\text{rel}}$ for
      fixed values of the c.m. momentum $K_{\text{c.m.}}$ (expressed in $fm^{-1}$)
      and two values of the
      angle $\Theta$ between ${\bf k}_{rel}$ and ${\bf K}_{\text{c.m.}}$, namely
      $\Theta = 90^0$ ({\it broken curves}) and $\Theta=0^0$ ({\it symbols}). In this Figure, and
      only in it,
      the continuous curves represent Eq. (\ref{factorization_pnB}) with $C_3^{pn}=2.0$.
      $^3$He  wave function from Ref. \cite{Kievsky:1992um,Kievsky:2010zr} and
      AV18 interaction \cite{Wiringa:1994wb}.
      (b): the same as in Fig.1(a) but for  $pp$ pairs.
      (c):  the $pn$ and $pp$ distributions corresponding to
       $K_{c.m.}=0$ in (a) and (b) and their sum.
      (d): the relative two-body momentum distributions
      $n^{N_1N_2}_{A}(k_{rel})= \int n^{N_1N_2}_{A}(\Vec{k}_{rel}, \Vec{K}_{c.m.})\,d\,\Vec{K}_{c.m.}$.
      In Fig. 1(c)
       the open and solid dots
    represent the   results from Argonne \cite{Wiringa:2013ala}.
    In this and the following Figures,
     unless  differently stated, the $pn$ and $pp$
      distributions are normalized to $ZN$ and $Z(Z-1)/2$, respectively.}
  \label{Fig1}
\end{figure*}
\begin{figure*}[!htp] 
  \centerline{\includegraphics[width=1.0\textwidth]{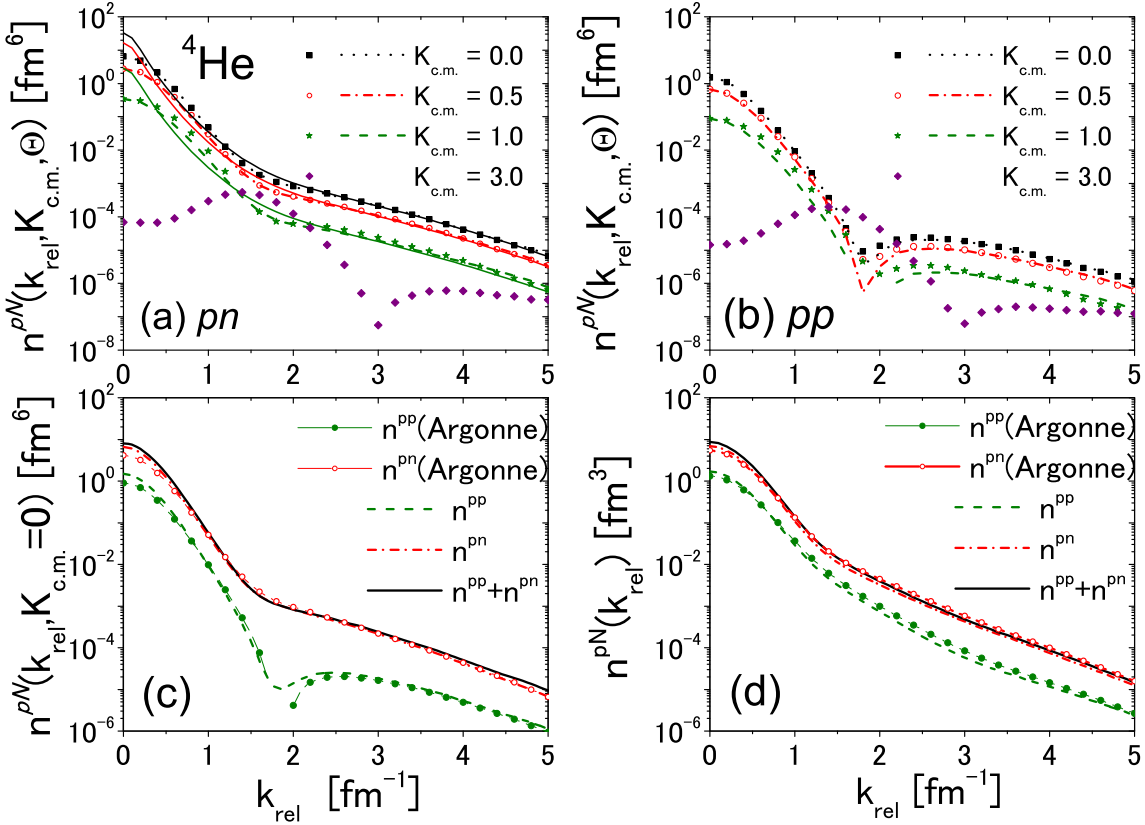}}
\vskip -1cm
  \caption{ (Color online) The same as in Fig. \ref{Fig1} but for $^4$He
    with  $C_4^{pn}=4.0$ in Fig.2(a). $^4$He wave function from
     Ref. \cite{Akaishi:1987}
      and
    AV8' interaction \cite{Pudliner:1997ck}. In Figs. 2(c) and 2(d)
       the open and solid dots
   denote the   results from Argonne \cite{Wiringa:2013ala}.}
  \label{Fig2}
\end{figure*}
\begin{figure}[!htp]  
\hspace{2.5cm}
  \centerline{\hspace{3.5cm}\includegraphics[width=0.58\textwidth]{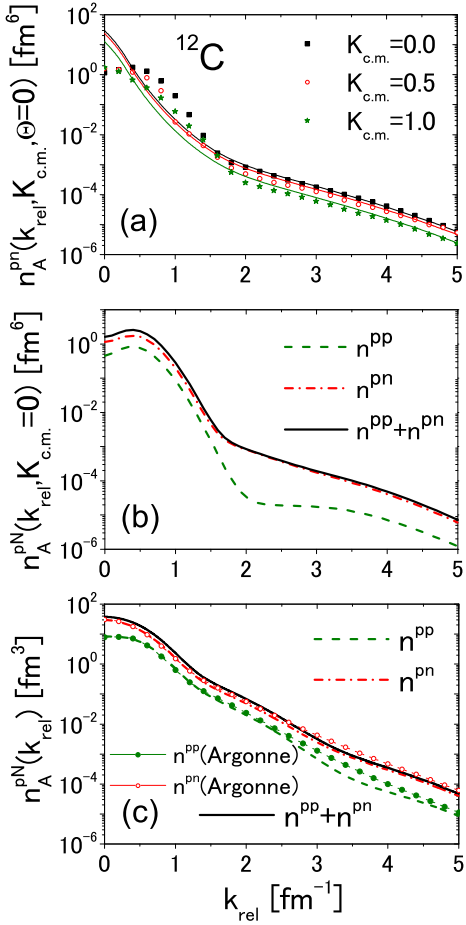}}
\vskip -1.0cm
  \caption{(Color online): the same as in Fig. \ref{Fig1}
    but for $^{12}$C with $C_{12}^{pn}=20.0$ in Fig.3(a).  $^{12}$C  wave
    function from Ref. \cite{Alvioli:2005cz} and AV8$^\prime$ interaction
    \cite{Pudliner:1997ck}. Note that in this Figure, as well as in Fig. \ref{Fig4}
    and \ref{Fig5} symbols in Figs. (a) correspond to $\Theta=0°$. In Fig. 3(c)
    the open and solid dots represent the results from Argonne \cite{Wiringa:2013ala}.}
  \label{Fig3}
\end{figure}
\begin{figure}[!htp] 
  \centerline{\hspace{3.5cm}\includegraphics[width=0.58\textwidth]{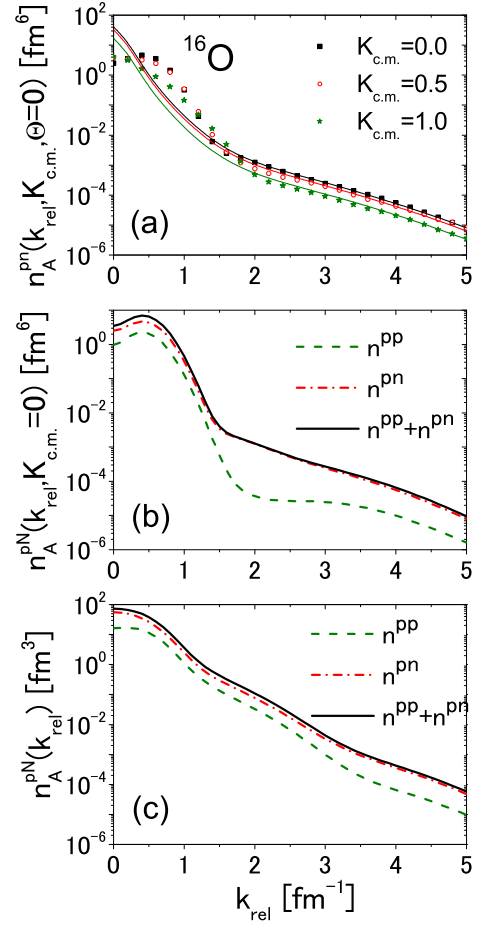}}
\vskip -1cm
  \caption{(Color online)   The same as in Fig. \ref{Fig1} but for $^{16}$O
    with $C_{16}^{pn}=24.0$ in Fig. 4(a). $^{16}$O  wave function from
    Ref. \cite{Alvioli:2005cz} and AV8' interaction \cite{Pudliner:1997ck}.}
  \label{Fig4}
\end{figure}
\begin{figure}[!htp] 
  \vskip -0.5cm
  \centerline{\hspace{3.5cm}\includegraphics[width=0.57\textwidth]{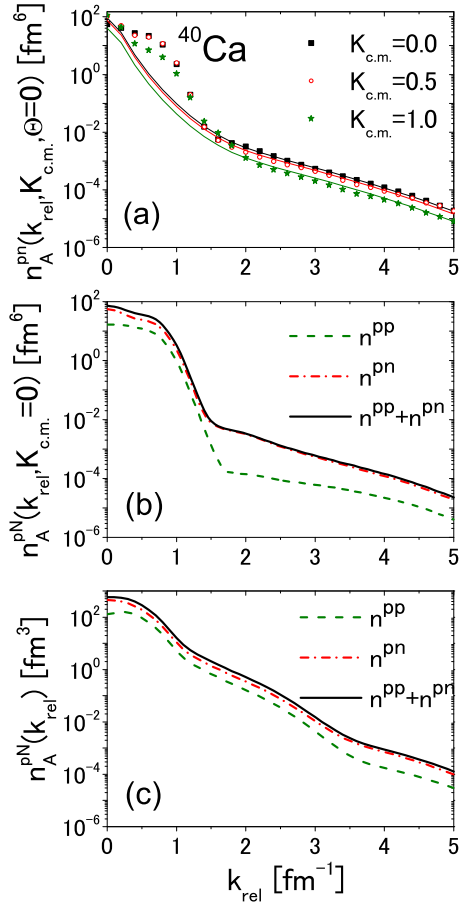}}
\vskip -0.5cm
  \caption{(Color online) The same as in Fig. \ref{Fig1} but for $^{40}$Ca
    with $C_{40}^{np}=60.0$ in Fig. 5(a). $^{40}$Ca  wave function from
    Ref. \cite{Alvioli:2005cz} and AV8' interaction \cite{Pudliner:1997ck}.}
  \label{Fig5}
\end{figure}
\begin{figure}[!htp] 
  \centerline{\includegraphics[width=0.5\textwidth]{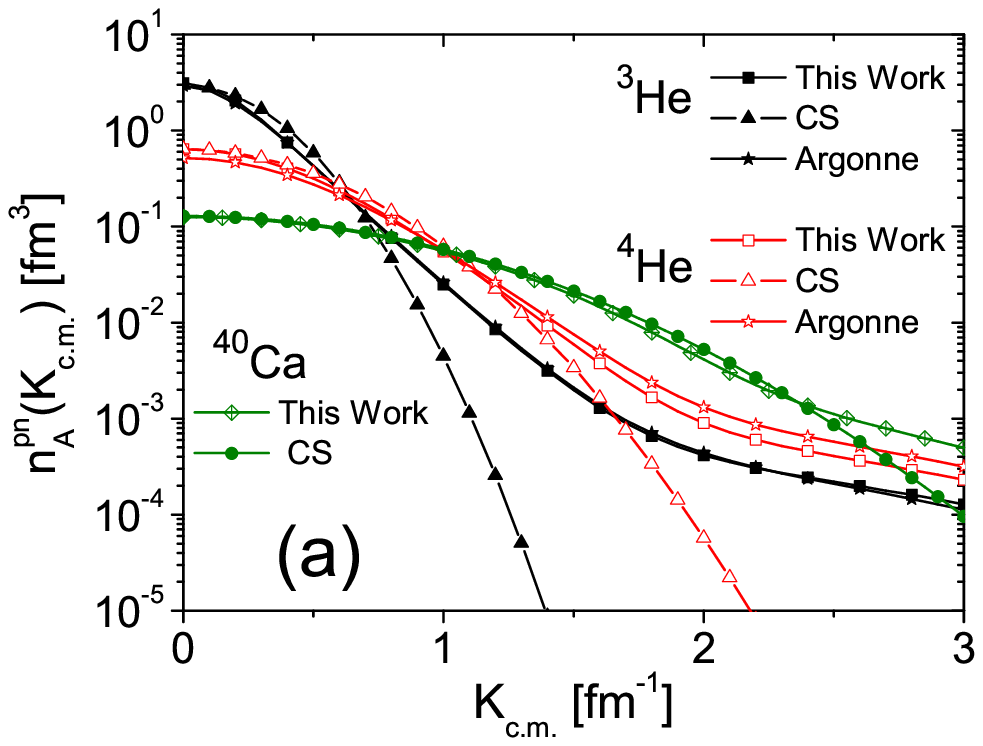}}
\vskip -1.0cm
  \centerline{\includegraphics[width=0.5\textwidth]{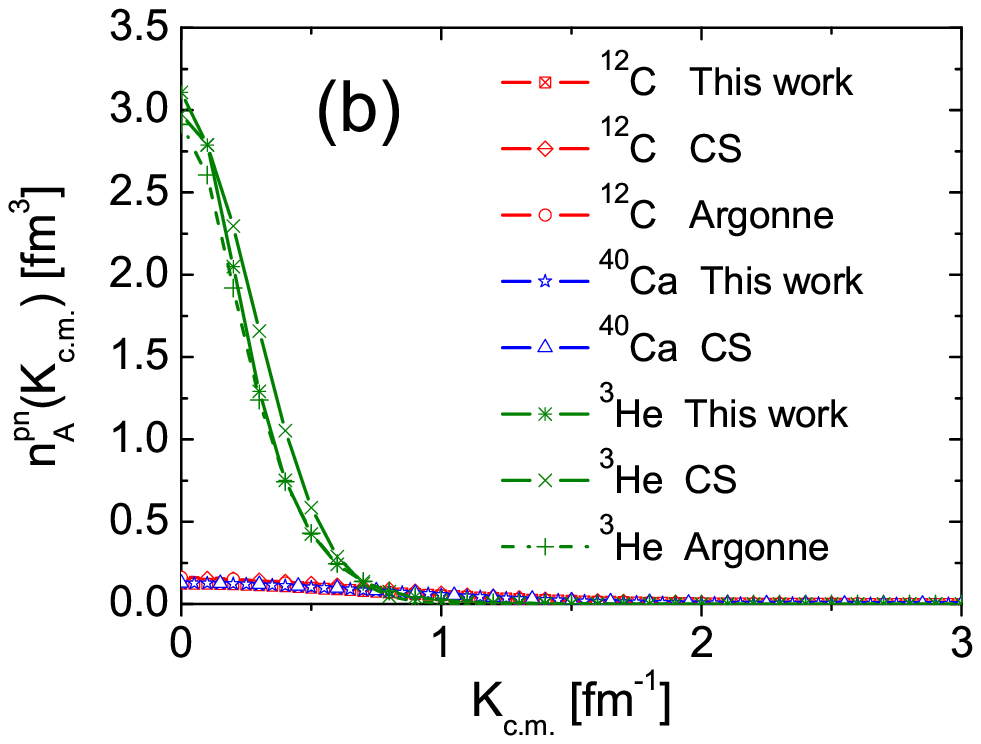}}
  \caption{(Color online)(a): The center-of-mass  momentum distribution
    $n_{\text{A}}^{pn}(K_{\text{c.m.}})= \int n_{A}^{pn}(\Vec{k}_{\text{rel}},\Vec{K}_{\text{c.m.}})\, d\,^3
    \Vec{k}_{\text{rel}}$
    (Eq. (\ref{n2_CM})) in $^3$He, $^4$He, $^{12}$C and $^{40}$Ca normalized to one, obtained in the present
    paper ({\it this work}), in Ref. \cite{CiofidegliAtti:1995qe} ({\it CS}) and in Ref. \cite{Wiringa:2013ala}
    (\textit{Argonne}). Note that a Gaussian distribution related to  the average value of the shell model
    kinetic energy \cite{CiofidegliAtti:1995qe} agrees very well with the many-body realistic  distribution
    up to $K_{c.m.} \simeq 1\,fm^{-1}$ except in the case of  $^{3}$H for which a shell model
      description has no meaning.
    (b):  the c.m momentum distributions of $^3$He, $^{12}$C and $^{40}$Ca on a linear scale.}
  \label{Fig6}
  \end{figure}
\begin{figure}[!hbp] 
  \vskip -0.5cm
  \centerline{\includegraphics[width=0.5\textwidth]{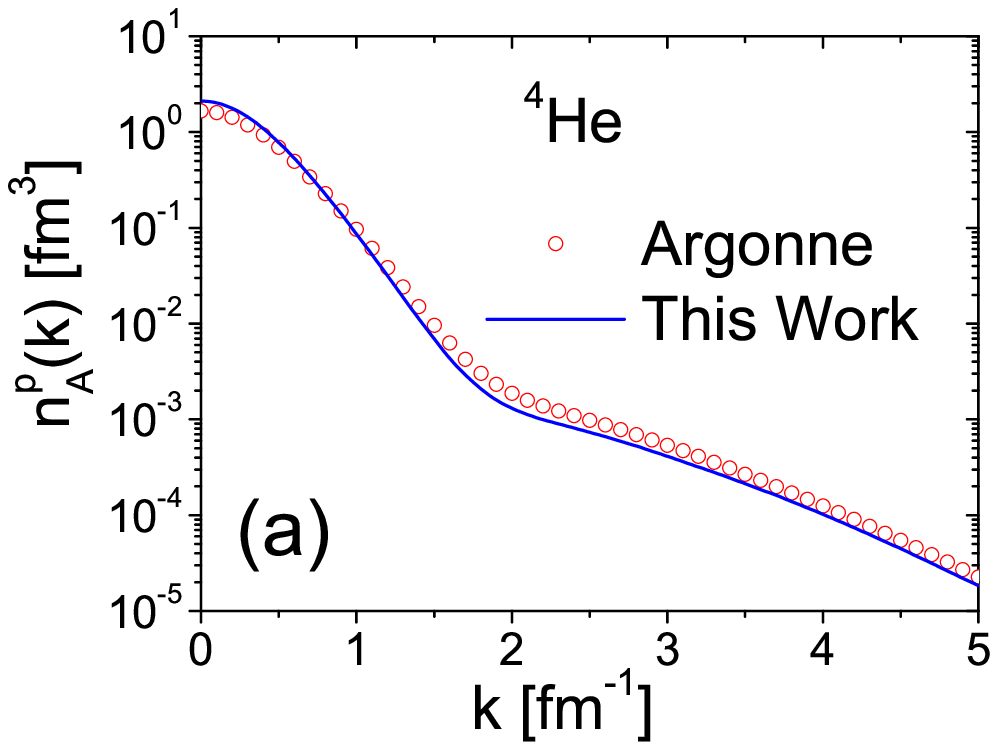}}
  \vskip -0.8cm
  \centerline{\includegraphics[width=0.5\textwidth]{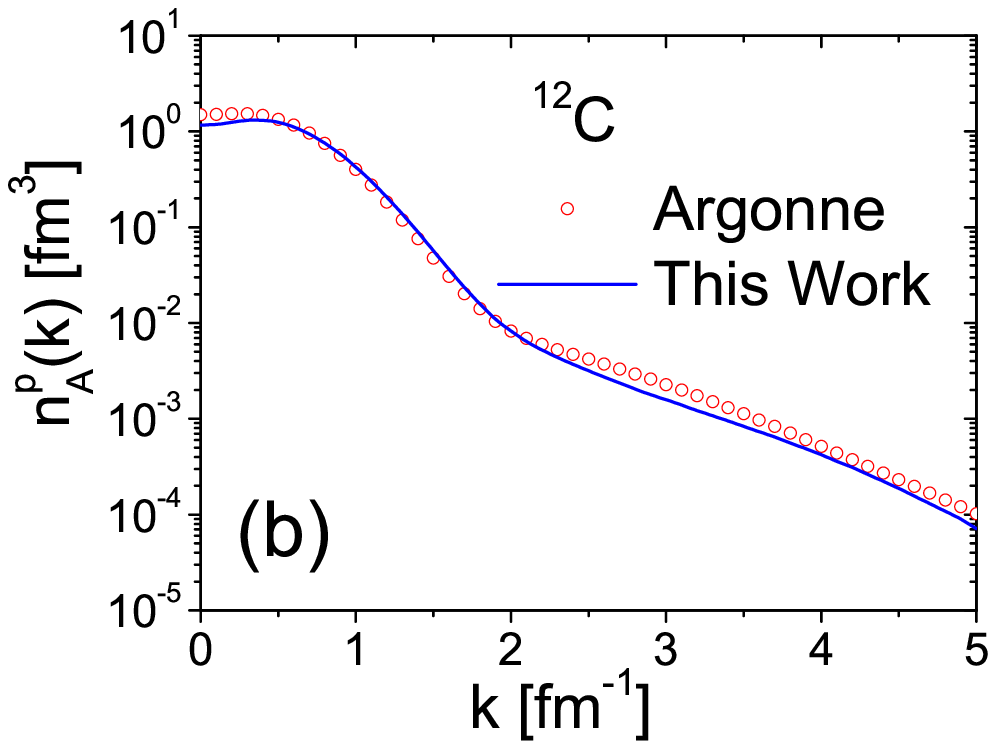}}
  \vskip -0.5cm
  \caption{(Color online)  (Comparison of the  one-nucleon momentum distributions
    of $^4$He and $^{12}C$ calculated in Ref. \cite{Alvioli:2012qa} (full line) and in  Ref.
    \cite{Wiringa:2013ala} (open dots), respectively. Normalization to the number of protons Z.}
  \label{Fig7}
  \vskip -0.5cm
\end{figure}

In this paper the number of protons and neutrons in nucleus A, will be denoted by Z and N, respectively, with A = Z + N. The two-body
momentum distributions of a pair of nucleons $N_1N_2$,  summed over spin (S)  and isospin (T) states, is given by
\begin{widetext}
\beqy n_A^{N_1N_2}(\Vec{k}_{1},\Vec{k}_{2})= \,\frac{1}{(2\pi)^6}\int d\Vec{r}_1\,d\Vec{r}_2\,d\Vec{r}_1^\prime\,d\Vec{r}_2^\prime\,
e^{i\,\Vec{k}_{1}\cdot\left(\Vec{r}_1-\Vec{r}_1^\prime\right)}\, e^{i\,\Vec{k}_{2}\cdot\left(\Vec{r}_2-\Vec{r}_2^\prime\right)}\,
\rho^{(2)}_{N_1N_2}(\Vec{r}_1,\Vec{r}_2;\Vec{r}_1^\prime,\Vec{r}_2^\prime)\,, \label{2bmomdis} \eeqy
\end{widetext}
where
\begin{widetext}
\beqy
\rho^{(2)}_{N_1N_2}(\Vec{r}_1,\Vec{r}_2;\Vec{r}_1^\prime,\Vec{r}_2^\prime)=
\int \psi_o^{*}(\Vec r_{1},\Vec r_{2},\Vec r_{3}...,\Vec{r}_A)\,
\psi_o(\Vec r_{1}^{\prime},\Vec r_{2}^{\prime},\Vec
r_{3},...,\Vec{r}_A)\,\delta\Big(\sum^A_{i=1}\Vec{r}_i\Big)\,
\prod\displaylimits_{i=3}^A d\Vec{r}_i\,,
\eeqy
\end{widetext}
is the two-body non-diagonal density matrix of nucleus $A$. The normalization of the proton, neutron and total distributions,
unless differently stated, is as follows \footnote{Note that in Ref. \cite{Alvioli:2011aa} the two-nucleon momentum
  distributions were normalized to one in the case of $^4$He, whereas in case of $^3$He it was normalized to the
  number of $pn$ and $pp$ pairs \textit{i.e.} two and one, respectively}
%
\beqy
&&\int n_A^{N_1N_2}(\Vec{k}_{1},\Vec{k}_{2})  d \Vec{k}_1\, d \Vec{k}_2
=\int \rho^{(2)}_{N_1N_2}(\Vec{r}_1,\Vec{r}_2) d \Vec{r}_1\, d \Vec{r}_2\nonumber\\
&&=\frac{Z(Z-1)}{2}\,{\Big |}_{N_1=N_2=p}\nonumber\\
&&=\frac{N(N-1)}{2}{\Big |}_{N_1=N_2=n}\nonumber\\
&&=ZN\,{\Big |}_{N_1=p,N_2=n} \label{Norma_partial}
\eeqy
%
with
\beqy
&&\sum_{N_1N_2}\int n_A^{N_1N_2}(\Vec{k}_{1},\Vec{k}_{2})  d \Vec{k}_1\, d \Vec{k}_2\nonumber\\
&&=\sum_{N_1N_2}\int\rho^{(2)}_{N_1N_2}(\Vec{r}_1,\Vec{r}_2) d \Vec{r}_1\, d \Vec{r}_2= \frac{A(A-1)}{2}\,. \label{Norma_total}
\eeqy
By introducing the relative and c.m. two-nucleon coordinates and momenta ($\Vec{r}=\Vec{r}_1-\Vec{r}_2$,
$\Vec{k}_{\text{rel}}=\left(\Vec{k}_1-\Vec{k}_2\right)/2$; $\Vec{R}=\left(\Vec{r}_1+\Vec{r}_2\right)/2$,
$\Vec{K}_{\text{c.m.}}=\,\Vec{k}_1+\Vec{k}_2 $), the two-nucleon momentum distribution can be rewritten
as follows \cite{Alvioli:2011aa}:
\beqy
&&n_A^{N_1N_2}(\Vec{k}_{\text{rel}},\Vec{K}_{\text{c.m.}})
=n_A^{N_1N_2}({k}_{\text{rel}},{K}_{\text{c.m.}},\Theta)\nonumber\\
&&=\frac{1}{(2\pi)^6}\int d\Vec{r}\,d\Vec{R}\,d\Vec{r}^\prime\,d\Vec{R}^\prime\,
e^{i\,\Vec{K}_{\text{c.m.}}\cdot\left(\Vec{R}-\Vec{R}^\prime\right)}\,\nonumber\\
&&\hspace{1cm}e^{i\,\Vec{k}_{\text{rel}}\cdot\left(\Vec{r}-\Vec{r}^\prime\right)}\,
\rho^{(2)}_{N_1N_2}(\Vec{r},\Vec{R};\Vec{r}^\prime,\Vec{R}^\prime),
\label{2brelc.m.}
\eeqy
where $|\Vec{k}_{\text{rel}}| \equiv k_{\text{rel}}$, $|\Vec{K}_{\text{c.m.}}| \equiv K_{\text{c.m.}}$  and $\Theta$
is the angle between $\Vec{k}_{\text{rel}}$ and ${\Vec K}_{\text{c.m.}}$. Of particular interest is the quantity
\beqy
&&n_A^{pn}({k}_{\text{rel}},{K}_{\text{c.m.}}=0)\nonumber\\
&&\hspace{1cm}=\frac{1}{(2\pi)^3}\int d\Vec{r}\,d\Vec{r}^\prime\,e^{i\,\Vec{k}_{\text{rel}}\cdot\left(\Vec{r}-\Vec{r}^\prime\right)}\,
\rho^{(2)}_{pn}(\Vec{r},\Vec{r}^\prime)\,, \label{back-to-back}
\eeqy
describing the spin and isospin  summed  relative momentum distributions of BB pairs,
 $\rho^{(2)}_{pn}(\Vec{r},\Vec{r}^\prime)$ being the c.m. integrated non-diagonal two-body density matrix.
  Relevant quantities are also the  $K_{c.m.}$- and
$k_{rel}$-integrated  momentum distributions, namely
\beqy
 n_A^{N_1N_2}({k}_{\text{rel}})= \int
\,n_A^{N_1N_2}(\Vec{k}_{\text{rel}},\Vec{K}_{\text{c.m.}})\,d\,\Vec{K}_{\text{c.m.}} \label{n2_rel} \eeqy
and
\beqy n_A^{N_1N_2}(K_{c.m.})=
\int \, n_A^{N_1N_2}(\Vec{k}_{\text{rel}},\Vec{K}_{\text{c.m.}}) d\,\Vec{k}_{\text{rel}}
 \label{n2_CM} \,.
\eeqy

Eqs. (\ref{back-to-back}), (\ref{n2_rel}) and  (\ref{n2_CM}) have been calculated in Refs. \cite{Feldmeier:2011qy,Roth:2010bm},
\cite{Schiavilla:2006xx}, \cite{Alvioli:2011aa} and \cite{Wiringa:2013ala}
for $^3$He and $^4$He, using {\it ab initio} wave functions and
in Ref. \cite{Wiringa:2013ala} for $^6$He, $^8$He, $^6$Li, $^7$Li, $^8$Li, $^9$Li, $^8$Be,
$^9$Be, $^{10}$Be, $^{10}$B and, preliminarily,  $^{12}$C, within the VMC approach.
In the this  paper we describe  new
results for various momentum distributions in  $^3$He, $^4$He, $^{12}$C, $^{16}$O and
$^{40}$Ca obtained, in the case of few-nucleon systems ($A=3, 4$), with {\it ab initio} wave functions, and, in the case
of nuclei with $A>4$, within a linked-cluster expansion
up to the order of four-body cluster contributions \cite{Alvioli:2005cz}.
Whenever possible  the  results  of our calculations of the
momentum distributions will be compared with the results
of the VMC approach of Ref. \cite{Wiringa:2013ala}
\footnote{In this paper we mainly discuss the spin-isospin summed momentum
  distributions of isoscalar nuclei  whereas  the spin-isospin dependent momentum distribution
  of non isoscalar nuclei will be the object of future investigations.}.
\section{Results of calculations and the universal properties of the correlated two-nucleon momentum distributions}\label{Sec:3}

\subsection{The two-nucleon momentum distribution in $^3$He and  isoscalar nuclei}
In Figs. \ref{Fig1}-\ref{Fig6} we show: (i) the $pn$ and $pp$ momentum distributions in $^{3}$He, $^{4}$He, $^{12}$C, $^{16}$O and
$^{40}$Ca nuclei, in particular, the full two-nucleon momentum distribution $n_A^{N_1N_2}(k_{rel}, K_{c.m.}, \Theta)$ (Eq.
(\ref{2brelc.m.})), (ii) the back-to-back momentum distributions (Eq. (\ref{back-to-back})), (iii) the relative momentum distributions, Eq.
(\ref{n2_rel}) and (iv)  the c.m. momentum distribution, (Eq. (\ref{n2_CM})). The results presented in these Figures have been obtained
using microscopic wave functions corresponding to the AV18 interaction \cite{Wiringa:1994wb} for $^2$H and $^3$He
\cite{Kievsky:1992um,Kievsky:2010zr} and the $AV8'$ interaction \cite{Pudliner:1997ck} for $^4$He \cite{Akaishi:1987} and complex nuclei
\cite{Alvioli:2005cz}. In order to  compare our results with the VMC results of Ref. \cite{Wiringa:2013ala} , whose wave functions are
calculated with 2N AV18+ 3N UX interaction,  we present in Fig. \ref{Fig7} the one-nucleon momentum distributions of $A=4$ and $A=12$
obtained  by the two approaches, even because both quantities will be used in what follows. Concerning our parameter-free  results,  let us
first of all stress that they are in
a general reasonable  agreement with the results of the VMC calculation \cite{Wiringa:2013ala}, although
in some regions of momenta (e.g. at $2.5 \lesssim k_{rel} \lesssim 3.5 \,fm^{-1}$) they can appreciably differ within  a 10-20 \%,
particularly in the case of the $pp$ relative momentum distribution
of $^{4}$He and $^{12}$C; the possible origin of such a disagreement,which
does not appear to be given  to the effects of the 3N force missing in our calculation \cite{Wiringa_PC},
is under investigation.
The obtained momentum distributions of both few-nucleon systems and complex nuclei exhibit
several universal features that can be summarized as follows:
\begin{enumerate}
\item as firstly pointed out in Ref. \cite{Schiavilla:2006xx} in the case of few-nucleon systems,
  when  $K_{\text{c.m.}}=0$,  the $pn$ and $pp$ momentum distributions do not
  appreciably differ at  small values of $k_{\text{rel}}$, with their ratio being
  closer to the ratio of the number of $pn$ to $pp$ pairs, whereas in the region
  $1.0\lesssim k_{\text{rel}} \lesssim 4.0$ fm$^{-1}$ the dominant role of tensor
  correlations makes the $pn$ distributions much larger than the $pp$ distribution,
  with the node exhibited by the latter filled up by the $D$ wave in the $pn$
  two-body density;
\item Figs. 1(a), (b) and 2(a), (b)  show that the momentum distribution $n_A^{NN}({k}_{\text{rel}},K_{\text{c.m.}}, \Theta)$,
  plotted {\it vs.} $k_{\text{rel}}$,
  decreases, at small and high values of $k_{rel}$, with increasing values of $K_{\text{c.m.}}$, whereas
  at intermediate values of $k_{\text{rel}}$ it increases with increasing values  of $K_{\text{c.m.}}$; this effect  is particularly
  relevant  for the $pp$ case where the dip occurring in the  $K_{c.m.}=0$ distribution  is totally washed out by the large $K_{c.m.}$
  components, resulting in a $K_{c.m.}$-integrated distribution totally
  different from the one corresponding to $K_{c.m.}=0$ ({\it cf.}); this effect seems to hold in the case of complex
  nuclei as well,
  as illustrated by the differences exhibited by Figures (b) and (c)  for A=12, 16, and 40.
\item  starting from a $K_{c.m.}$-dependent  value of the relative momentum $k_{rel}$,  to
  be denoted $k_{\text{rel}}^{-}(K_\text{c.m.})$, the $pn$ two-nucleon momentum distributions become
  to a large extent $\Theta$ independent ,
  with the value of $k_{\text{rel}}^{-}(K_\text{c.m.})$ increasing
  with $K_\text{c.m.}$, according to the following relation
  \beqy
  k_{\text{rel}}^-(K_\text{c.m.})=a_1+ f(K_{\text{c.m.}}) \equiv k_{\text{rel}}^-;
  \label{k_rel_fac}
  \eeqy
  that can be defined with    $a_1 \simeq 1.5\,fm^{-1}$ ({\it cf}  Figs. \ref{Fig1}-\ref{Fig5})
  and $f(K_{\text{c.m.}}) =K_{c.m.}$;
  $\Theta$ independence,
  firstly stressed in Ref. \cite{Alvioli:2012qa} and verified in a wide range of angles, implies that for $k_ {\text{rel}}>k_{\text{rel}}^-$ the two-nucleon
  momentum distribution factorizes, \textit{i.e.} $n_A^{N_1N_2}({k}_{\text{rel}}, {K}_{\text{c.m.}},\Theta)
  \propto n_{\text{rel}}^{N_1N_2} ({k}_{\text{rel}})n_{\text{c.m.}}^{N_1N_2}({K}_{\text{c.m.}})$. In the region of
  factorization, defined  by $k_{{\text{rel}}}\gtrsim k_{{\text{rel}}}^{-}$ and
  $K_{{\text{c.m.}}}\lesssim 1\,fm^{-1}$, the momentum distribution for $pn$ pairs can
  be approximated
  as follows:
  \beqy
  &&n_A^{pn(fact)}({k}_{\text{rel}},{K}_{\text{c.m.}})\nonumber\\
  &&\hspace{1cm}\simeq\frac{n_A^{pn}({k}_{\text{rel}},{K}_{\text{c.m.}}=0)}
       {n_{\text{c.m.}}^{pn}(K_{\text{c.m.}}=0)}
       \,n_{\text{c.m.}}^{pn}(K_{\text{c.m.}})\nonumber\\
       \label{factorization_pnB}
       &&\hspace{1cm}\simeq C_A^{pn}
       n_D({k}_{\text{rel}}) n_{\text{c.m.}}^{pn}(K_{\text{c.m.}})\,.
  \eeqy
  Here $n_D(k_{rel})$ is the deuteron momentum distribution, $n_{c.m.}^{pn}(K_{c.m.})$
  the c.m. momentum distribution of the correlated pair in the region of factorization and $C_A^{pn}$  an A-dependent
  constant, whose  value and physical meaning will  be discussed in the next Subsection.
  As for the $pp$ momentum distribution,  it appears that it also factorizes but starting at
  a value of  the relative momentum higher than  $k_{{\text{rel}}}(K_{c.m.})^-$;
  one has anyway
  \beqy
  &&n_A^{pp(fact)}({k}_{\text{rel}},{K}_{\text{c.m.}})\nonumber\\
  &&\hspace{1cm}\simeq\frac{n_A^{pp}({k}_{\text{rel}},{K}_{\text{c.m.}}=0)}
      {n_{\text{c.m.}}^{pp}(K_{\text{c.m.}}=0)}
      \,n_{\text{c.m.}}^{pp}(K_{\text{c.m.}})\nonumber\\
      &&\hspace{1cm}\simeq C_A^{pp}
      n_{rel}^{pp}({k}_{\text{rel}}) n_{\text{c.m.}}^{pp}(K_{\text{c.m.}})\,\,.
      \label{factorization_pp}
  \eeqy
\begin{figure}[!hbp] 
  \includegraphics[width=0.5\textwidth]{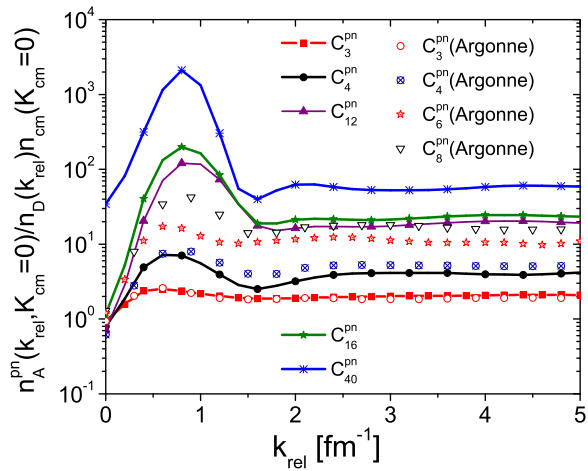}
  \vskip -0.5cm
  \caption{(Color online) Determining the constant $C_A^{pn}$ by a plot of  Eq. (\ref{CA})
  {\it vs} $k_{rel}$ and fixed value of $K_{c.m.}=0$. The constant value of Eq. (\ref{CA})
  determines the  value of  $C_A^{pn}$. For $^3$He, $^4$He, $^6$Li and $^8$Be the results
  obtained with the Argonne momentum distributions  \cite{Wiringa:2013ala}
  are also shown.}
  \label{Fig8}
\end{figure}
  where, unlike the $pn$ case,   the  momentum distribution
  $n_{rel}^{pp}({k}_{\text{rel}})$
  is, at the moment, not defined in terms of a  $pp$ system.
  Eqs. (\ref{factorization_pnB}) and (\ref{factorization_pp})
  describe a property  exhibited in Figs. \ref{Fig1} and \ref{Fig2}
  (and common to  any value of A), namely  that at high values of
$k_{rel}>k_{rel}^-(K_{c.m.})$  the $pN$ momentum
  distributions  differ only by their magnitudes, which are governed by
   $n_{c.m.}^{pN}$
   \footnote{Note that $n_A^{N_1N_2}(K_{c.m.})$ (Eq. (\ref{n2_CM}))
   includes all c.m. momentum components,
   whereas $n_{c.m.}^{N_1N_2}(K_{c.m.})$ has to describe only
   the low-momentum part
   ($K_{c.m.} \lesssim 1-1.5\,fm^{-1}$)  of the c.m. motion};
\item at high values of the relative and c.m. momenta,  more
  than two particles can be locally correlated, producing a strong dependence upon the angle
   $\Theta$
  and, correspondingly,  the violation  of factorization, as  shown in Fig. \ref{Fig1}
 in the case of $K_{\text{c.m.}}= 3 \, fm^{-1}$; moreover, it can be seen ({\it cf} Fig. \ref{Fig1}(b)
 and \ref{Fig2}(b)) that the behavior of $n_A^{pp}$ in the region around $k_{rel} \simeq 2\, fm^{-1}$ is strongly affected by the high $K_{c.m.}$
 momentum components;
\item in Ref. \cite{CiofidegliAtti:1995qe} the low momentum part ($K_{\text{c.m.}} \lesssim 1.0$ fm$^{-1}$)
  of the {\it c.m.} momentum distribution has been  described by a gaussian function normalized to one, namely,
  $n_{c.m.}^A(K_{c.m.}) = (\alpha_A/\pi)^{3/2}exp (-\alpha_A\,K_{c.m.}^2)$, with the values of $\alpha_A$
  obtained from the average value of the shell model  kinetic energy $<T>_{SM}$, as follows
  $\alpha_A=\frac{3(A-1)}{4 m_N (A-2)<T>_{SM}}$. It can be seen from Fig. \ref{Fig6} that, apart
  from the case of $^3$He, for which  a shell-model   description is meaningless, the Gaussian model
  of Ref. \cite{CiofidegliAtti:1995qe} nicely approximates the many-body result in the region of
  $K_{cm}\lesssim 1\, fm^{-1}$. The values of $\alpha_A$ for $^4$He and $^{12}$C obtained in
  Ref. \cite{CiofidegliAtti:1995qe} also agree with the experimental data
  \cite{Piasetzky:2006ai,Shneor:2007tu,Korover:2014dma}, to be discussed in Section \ref{Sec:5}.
\end{enumerate}
\begin{table*}[!ht] 
  \normalsize
  \begin{center}
    {\renewcommand\arraystretch{1.3}
      \begin{tabular}{|c|c|c|c|c|c|c|c|}
        \hline
        { }$^2$H{ }&
        $^3$He&
        $^4$He&
        $^6$Li&
        $^{8}$Be&
        $^{12}$C&
        $^{16}$O&
        $^{40}$Ca\\
        \hline
        1.0&
        2.0 $\pm$ 0.1&
        4.0 $\pm$ 0.1&
        --&
        --&
        20 $\pm$ 1.6&
        24 $\pm$ 1.8&
        60 $\pm$ 4.0\\
        1.0&
        (2.0 $\pm$ 0.1)&
        (5.0$\pm$ 0.1)&
        (11.1 $\pm$ 1.3)&
        (16.5 $\pm$ 1.5)&
        (--)&
        (--)&
        (--)\\
        \hline
        \hline
    \end{tabular}}
  \end{center}
  \caption{The values of the constant $C_A^{pn}$ (Eq. (\ref{CA}))
    extracted from Fig. \ref{Fig8}, with error determined according to the
    following expression:
    $C_A^{pn}=\frac{(C_A^{pn})^{Max}+(C_A^{pn})^{Min}}{2} \pm
    \frac{(C_A^{pn})^{Max}-(C_A^{pn})^{Min}}{2}$,
     where $(C_A^{pn})^{Max}$ and $(C_A^{pn})^{Min}$ are determined in the
     region of $k_{rel}\geq 3.0 fm^{-1}$. The values in
    brackets have been
    obtained using the VMC wave function of Ref. \cite{Wiringa:2013ala}}
  \label{Table1}
\end{table*}

%
\subsection{The meaning and the numerical values  values of the quantity $C_A^{pn}$}
In what follows we will discuss in detail   the behavior of the $pn$ momentum distributions in the correlation region, in particular  the
meaning and the numerical value  of the constant $C_A^{pn}$ appearing in Eq. (\ref{factorization_pnB}).  This is because we would like to
compare the short-range behavior  of a bound $pn$ pair, \ie the deuteron, with the behavior of a $pn$ pair in the nuclear medium. The
factorized form (Eq. (\ref{factorization_pnB})) describes 2N SRCd configurations when the relative momentum of the pair is much larger than
the c.m. momentum. Since for isoscalar nuclei $n_{\text{c.m.}}^{pn}\simeq n_{\text{c.m.}}^{pp}$, the A-dependence of $ n_A^{pn(fact)}$ is
given only by the A-dependence of both the constant $C_A^{pn}$  and by the c.m. momentum distribution $n_{c.m.}^{pn}$, with the  former
determining   the amplitude of $n_A^{pn(fact)}({k}_{\text{rel}},{K}_{\text{c.m.}}=0)$ and the latter its damping
with increasing values of $K_{c.m.}$, as it clearly appears from Figs. \ref{Fig1}-\ref{Fig5}, where can indeed  be seen that the decrease
of $n_{A}^{pn}({k}_{\text{rel}},{K}_{\text{c.m.}})$ at $k_{rel}>k_{rel}^-$ exactly follows the rate of decrease of
$n_{A}^{pn}(K_{\text{c.m.}})$ shown in Fig. \ref{Fig6}, whose low $K_{\text{c.m.}}$ distribution coincides with
$n_{c.m.}^{pn}(K_{\text{c.m.}})$.
 Therefore  it can be concluded that  $C_A^{pn}$: (i) is independent of $k_{rel}$
and $K_{c.m.}$, \ie it is a quantity depending only upon the value of  $A$, (ii) it is not a free and adjustable parameter, but a quantity
resulting  from  {\it ab initio} many-body calculations of the momentum distributions, since, (iii)
 it is defined   in terms
of  the magnitude of $n_A^{pn}(k_{rel}, K_{c.m.}=0)$ at $k_{rel}\gtrsim k_{rel}^-$,  the deuteron momentum distribution, and, eventually,
by the c.m. momentum distribution of the pair, \ie by  quantities  resulting from  many-body calculations and from the factorization
property of the momentum distributions. To sum up,  the value of $C_A^{pn}$ is given by  the following relation
\beqy
&&\lim_{k_{rel}> k_{rel}^-}{\frac{n_A^{pn}(k_{rel},K_{c.m.}=0)}{n_{c.m.}^{pn}(K_{c.m.}=0)n_D(k_{rel})}}\nonumber\\
&&\hspace{1cm}=Const \equiv C_A^{pn}
\label{CA}
\eeqy
The validity of Eq. (\ref{CA}) and the determination of the value of $C_A^{pn}$ are illustrated in Fig. \ref{Fig8}.
It can be seen that at low values of the relative momentum ($k_{rel} \lesssim 1.5 \, fm^{-1})$ the ratio Eq. (\ref{CA})
exhibits  a strong dependence upon $k_{rel}$,  reflecting the A-dependent mean-field structure  whereas, starting from
$k_{rel} \simeq 2-2.5 \, fm^{-1}$,  a constant behavior  is observed for all values of $A$ that have been considered;
in particular, in the case of $A=3$ and  $A=4$, for which  accurate   wave functions have been used, the consistency
with a constant value is very good, whereas for complex nuclei, which are more sensitive to the many-body approximations,
the error on the determination of the value of  $C_A^{pn}$ is higher. The obtained values of $C_A^{pn}$ are listed in Table
\ref{Table1}, where the values obtained with the VMC results of Ref. \cite{Wiringa:2013ala} are also shown in brackets.
The difference in the value of $C_{{A=4}}^{pn}$ between ours and the VMC approaches
  could be attributed to the different Hamiltonian (V8' NN interaction in our case
  and AV18 in VMC method) and to the different variational wave functions,
  whereas in the case
  of heavier nuclei possible effects from the omitted terms of the cluster expansion
  should also be considered. All of these possibilities
are under investigation.  Nonetheless the results of both approaches exhibit the same  A-dependency, \ie an increase of the value of
$C_A^{pn}$ with the value of $A$, which confirms the factorization property of the momentum distribution and that can  be
explained with the very physical meaning of $C_A^{pn}$. As a matter of fact Eq. (\ref{factorization_pnB})  provides the physical meaning of
the constant $C_A^{pn}$, namely in the factorization region one obtains
\beqy
&&n_{pn}^{SRC,BB}(K_{c.m.}=0)\nonumber\\
&&=\int_{{k_{rel}^-}=1.5}^{\infty} d\,\Vec{k}_{rel}\int_{0}^{\infty}
    {n_A^{pn}({\bf k}_{\text{rel}},{\bf K}_{\text{c.m.}})}\delta({\bf K}_{c.m.})\,d\,\Vec{K}_{c.m.}\nonumber\\
    &&\simeq C_A^{pn}\,n_{c.m.}^{pn}(K_{c.m}=0)\,4\pi\int_{k_{rel}^{-}=1.5}^{\infty} n_D(k_{rel}) k_{rel}^2\,dk_{rel}\nonumber\\
    \label{NNBB}
\eeqy
 which represents the momentum distribution  of back-to back (BB) nucleons integrated
 in the region of relative momentum $k_{rel} \geq 1.5 \,fm^{-1}$. Thus $C_A^{pn}$
 represent a measure of the number of SRCd $pn$ pairs with c.m. momentum
 distribution $n_{c.m.}^{pn}(K_{c.m.}=0)$,  \ie the number of  deuteron-like pairs. At the same
 time the equation
 \beqy
&&\hspace{-0.4cm}N_{pn}^{SRC}=\int_{0}^{K_{c.m.}^{max}}
   d\,\Vec{K}_{c.m.}\int_{k_{rel}(K_{c.m.})}^{\infty}{n_A^{pn}}({\bf k}_{\text{rel}},{\bf K}_{\text{c.m.}})d\,\Vec{k}_{rel}
   \nonumber\\
 &&\simeq C_A^{pn}\,(4\pi)^{2}\int_{0}^{K_{c.m.}^{max}}
 \,n_{c.m.}^{pn}(K_{c.m.})\,{K}_{c.m.}^2\,d\,{K}_{c.m.}\nonumber\\
 &&\hspace{1cm}\int_{k_{rel}^-(K_{c.m.})}^{\infty} \,n_D(k_{rel})\,{k}_{rel}^2\,d\,{k}_{rel}\,,
 \label{total_SRC}
 \eeqy
represents the the number of SRCd $pn$ pairs in  the  entire  two-nucleon SRC region, characterized by $K_{c.m}^{max}\lesssim 1 \sim
1.5\, fm^{-1}$ and $k_{rel}^-\gtrsim 1.5\, fm^{-1}$ \footnote{Following the original suggestion  of Ref. \cite{Frankfurt:1988nt} we also
adopt here the region
$k_{rel}^-\gtrsim 1.5\, fm^{-1}$ as the {\it SRC region}, although, more correctly,
as it appears from the results of many-body calculations, the {\it SRC region} starts
from $k_{rel}^-\gtrsim2\,fm^{-1}$.}.
\section{The factorization property of the nuclear wave function
  and the high momentum behavior of the momentum distributions} \label{Sec:4}
\subsection{SRCs as a result of wave function factorization}
\begin{figure*}[!htp] 
  \includegraphics[width=0.9\textwidth]{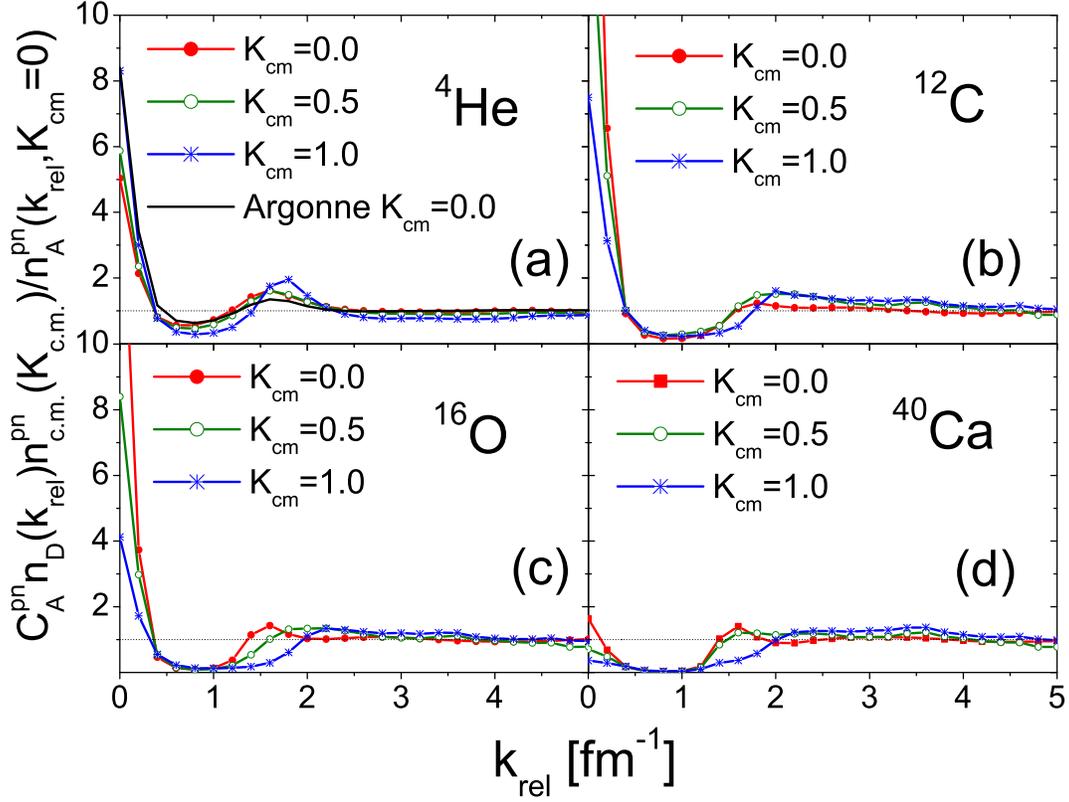}
  \vskip -1.0cm
  \caption{ (Color online)
  The ratio (Eq. (\ref{ratio_fac__ex}))  between the factorized distributions
    (Eq. (\ref{factorization_pnB})) and the exact  ones ($\Theta=0^o$) for $^4$He, $^{12}$C, $^{16}$O and
     $^{40}$Ca in correspondence of  $K_{c.m.}=0, \, 0.5, \, 1 \,fm^{-1}$. For $^4$He the results obtained
     with the Argonne momentum distributions \cite{Wiringa:2013ala} are  shown by the full line.}
  \label{Fig10}
\end{figure*}
\begin{figure}[!hbp] 
  \vskip -1.2cm
  \includegraphics[width=0.5\textwidth]{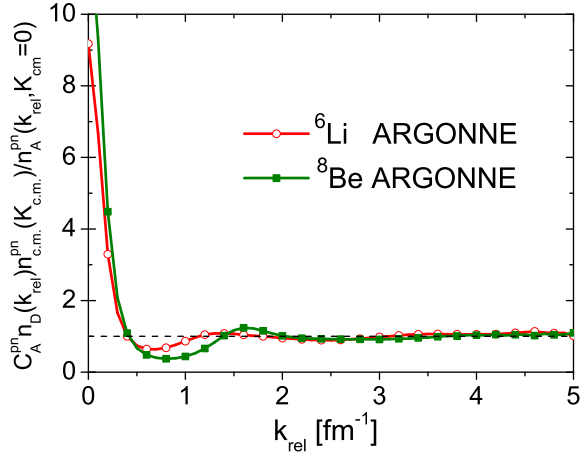}
  \vskip -0.5cm
  \caption{(Color online)
  The ratio (Eq. (\ref{ratio_fac__ex})) between the factorized distributions,
  Eq. (\ref{factorization_pnB}), and the exact ones for $^6$Li and $^{8}$Be,
  corresponding to
  $K_{c.m.}=0$, obtained with the Argonne VMC wave functions \cite{Wiringa:2013ala}.}
  \label{Fig11}
  \vskip -0.2cm
\end{figure}%

It has been  demonstrated that the the momentum distributions of nuclei in the region
of SRCs are governed by the {\it factorization property} of the nuclear wave function at
short inter-nucleon distances, described by the following relation
\beqy
&&\lim_{r_{ij}\rightarrow 0}  \Psi_0(\{\Vec r\}_A)\nonumber\\
&&\simeq \mathcal{\hat A}
\Big\{\chi_o(\Vec R_{ij})
\sum_{n,f_{A-2}}a_{o,n,f_{A-2}}\Big[ \Phi_n(\Vec x_{ij},\Vec r_{ij})\nonumber\phantom{\Big]}\phantom{\Big\}}\\
  &&\phantom{\Big\{}\phantom{\Big[}\oplus\Psi_{f_{A-2}}(\{\Vec x \}_{A-2}, \{\Vec r \}_{A-2})\Big] \Big\},
\label{wf_fact}
\eeqy
where: {i) $\{\Vec r\}_A$ and $\{\Vec r\}_{A-2}$ denote the set of radial coordinates of nuclei $A$ and $A-2$,
respectively; (ii)   $\Vec r_{ij}$ and $\Vec R_{ij}$ are the relative and c.m. coordinate of the nucleon pair $ij$, described by the
relative  wave function $\Phi_n$ and the c.m. wave function $\chi_o$  in $0s$ state; iii) $\{\Vec x \}_{A-2}$ and  $\Vec x_{ij}$ denote the
set of  spin-isospin coordinates
 of the nucleus  $(A-2)$ and the pair $(ij)$.
Eq. (\ref{wf_fact}) has been introduced  in \cite{CiofidegliAtti:1995qe}
demonstrating  that the SRCd nuclear two-nucleon momentum distribution factorizes
into the vector-coupled product of the  relative and c.m. momentum distribution of a NN pair.
In particular, in Ref. \cite{CiofidegliAtti:2010xv} the factorization property of the nuclear wave
function has been shown to hold in the case of {\it ab} initio wave functions of few-nucleon
systems, showing that the momentum-space wave function of $^3$He and $^4$He factorize
in the region of {\it high } ($k_{rel}^{-}\gtrsim 2 fm^{-1}$) relative momenta coupled
to \textit{low}  c.m. momenta,  ($K_{c.m.}\lesssim 1.0 fm^{-1}$), whereas  at higher values
of $K_{c.m.}$  factorization still occurs but  starting at increasing  values of $k_{rel}$; such
a behavior  indeed  appears  in Figs. \ref{Fig1}(a)-\ref{Fig5}(a), both  in the case of few-nucleon
systems and complex nuclei. Finally, in Ref \cite{Baldo:1900zz}, the factorization property of the
wave function and momentum distribution have also been shown to occur in case of nuclear matter
treated within the Brueckner-Bethe-Goldstone approach.
In order to provide new evidence about the validity of the  factorization property, we show in Figs. \ref{Fig10} and \ref{Fig11} the ratio
of the factorized momentum distribution of a $pn$ pair, Eq. (\ref{factorization_pnB}) to the exact momentum distribution $n_A^{pn}
({k_{rel},K_{cm},\Theta})$, \textit{i.e} the quantity
\beqy
R_{fact/exact}^{pn}=\frac{C_A^{pn}n_D(k_{rel})n^{pn}_{c.m.}(K_{c.m.})}{n_A^{pn}({k_{rel},K_{cm},
    \Theta})}
\label{ratio_fac__ex}
\eeqy
plotted on a linear scale. It can be seen that, independently  of the nuclear mass and the values of $K_{c.m.}$, the ratio exhibits, at
$k_{rel}^-\gtrsim 2 \, fm^{-1}$, a  constant value equal to one. The scaling to one is perfect  for  $A= 4\,, 6\,, 8$  nuclei for which
{\it ab initio} VMC momentum distributions  have been  used, whereas it presents small oscillations for complex nuclei, a behavior that
should be attributed to the approximations which have been used to solve the many-body problem.

\subsection{Wave function factorization and the relation between the relative momentum distribution of $pn$
  pairs in nuclei and the deuteron momentum distributions}
In Fig.\ref{Fig12}  the  two-nucleon  momentum distributions of $pn$ pairs  in nuclei is compared with
 the deuteron momentum distribution.
\begin{figure*}[!htp] 
  \centerline{\hspace{0.7cm}\includegraphics[width=0.50\textwidth]{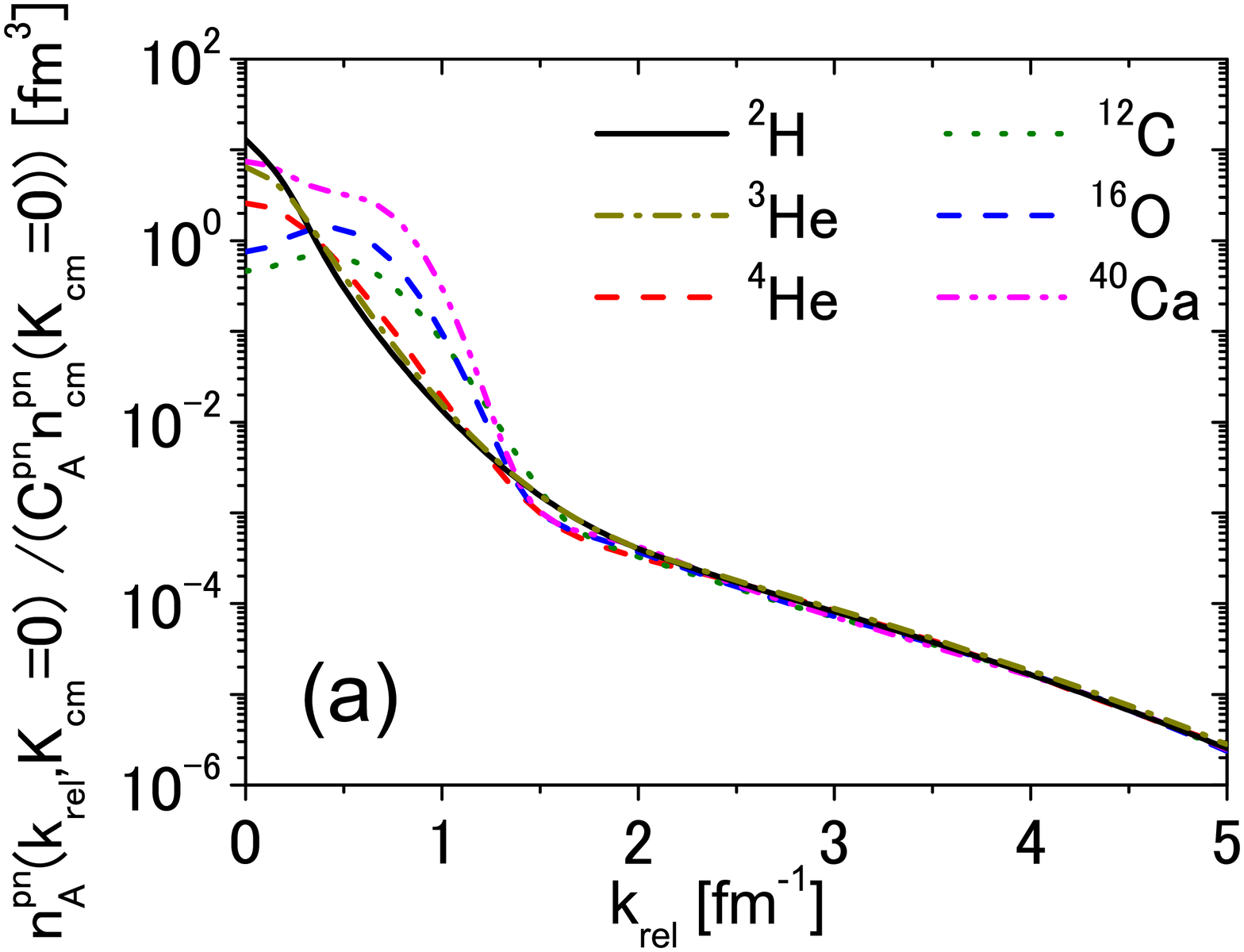}\hspace{-1.0cm}
    \includegraphics[width=0.50\textwidth]{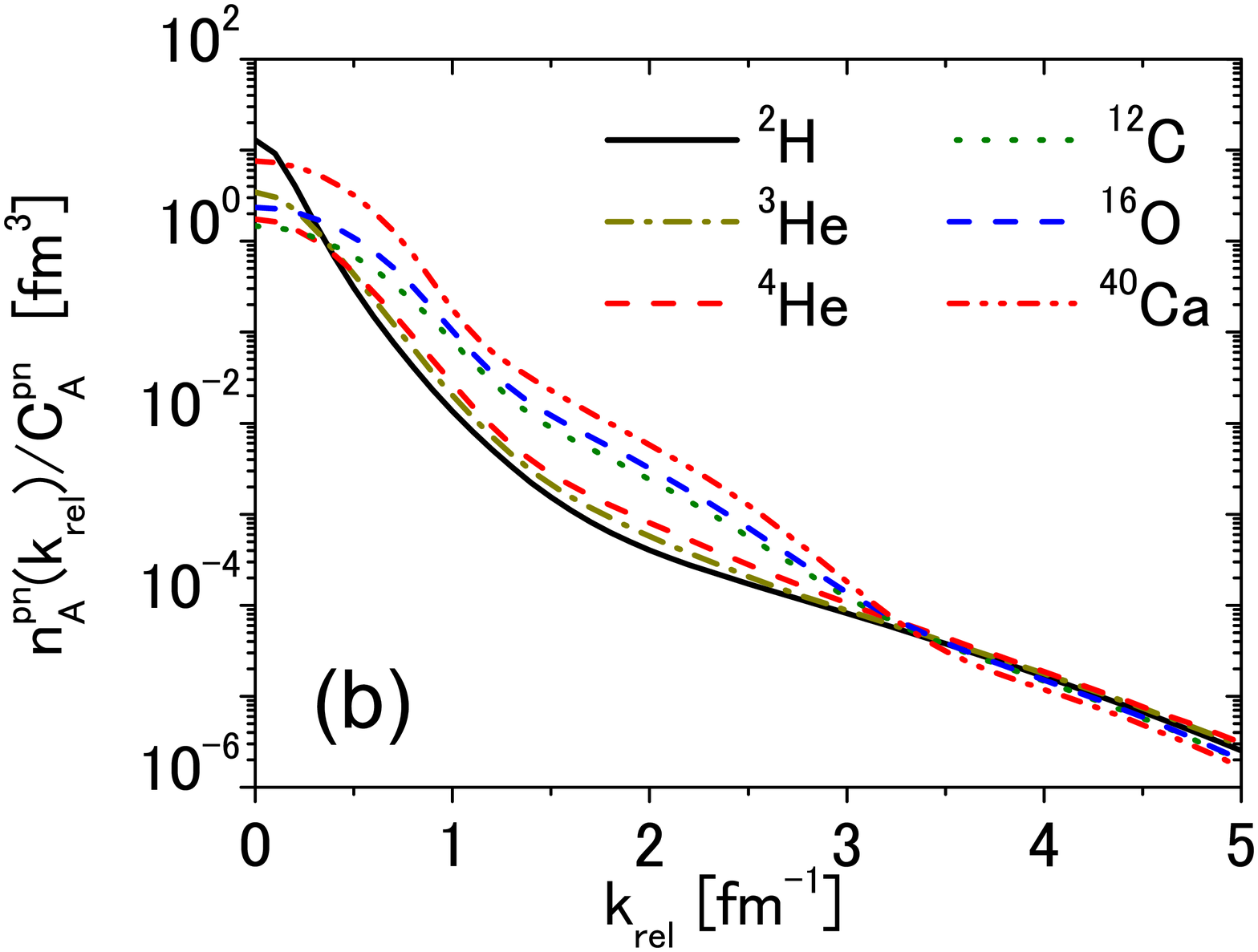}}
  \vskip -0.5cm
  \centerline{\hspace{0.10cm}\includegraphics[width=0.50\textwidth]{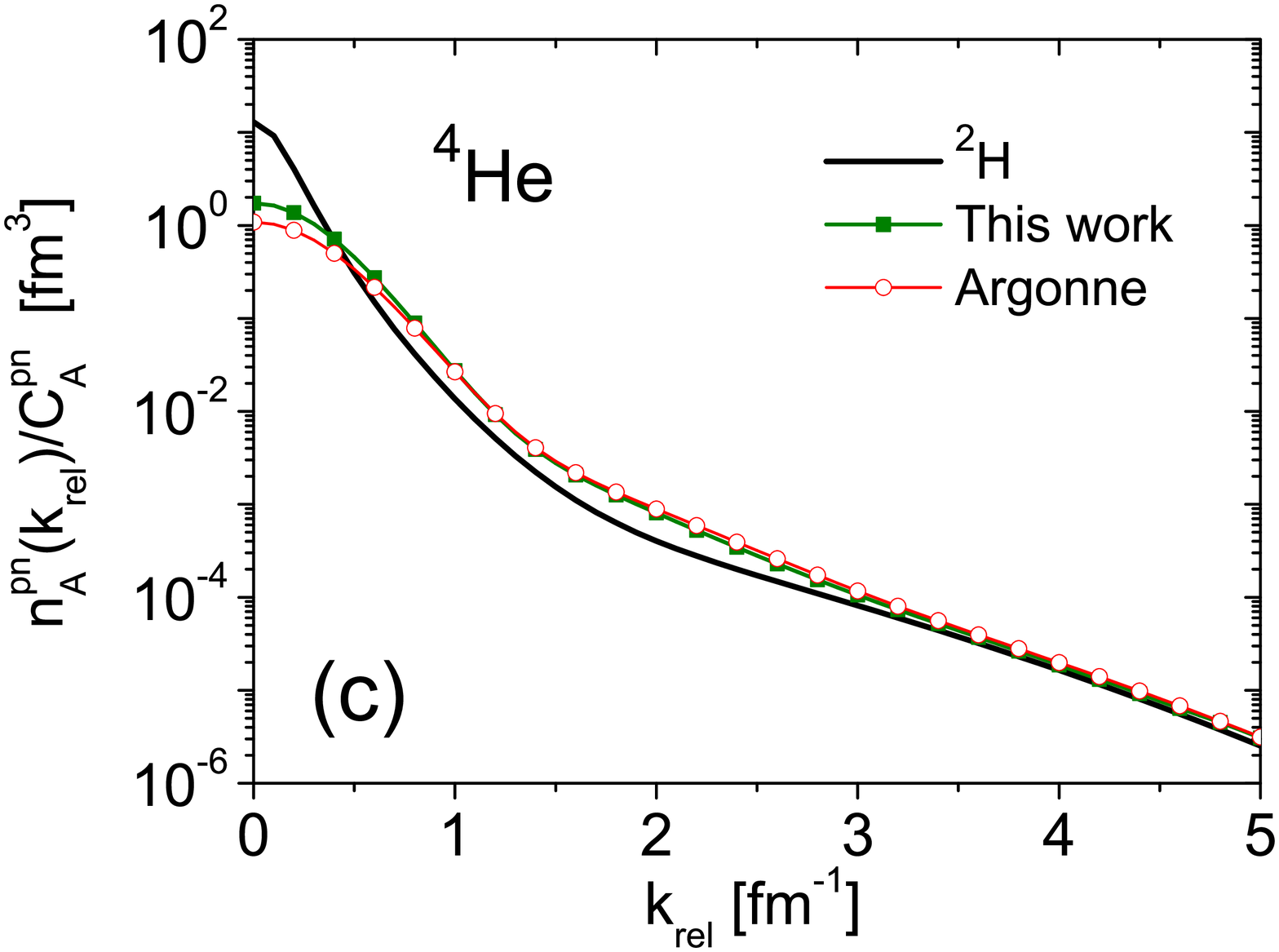}\hspace{-1.0cm}
    \includegraphics[width=0.50\textwidth]{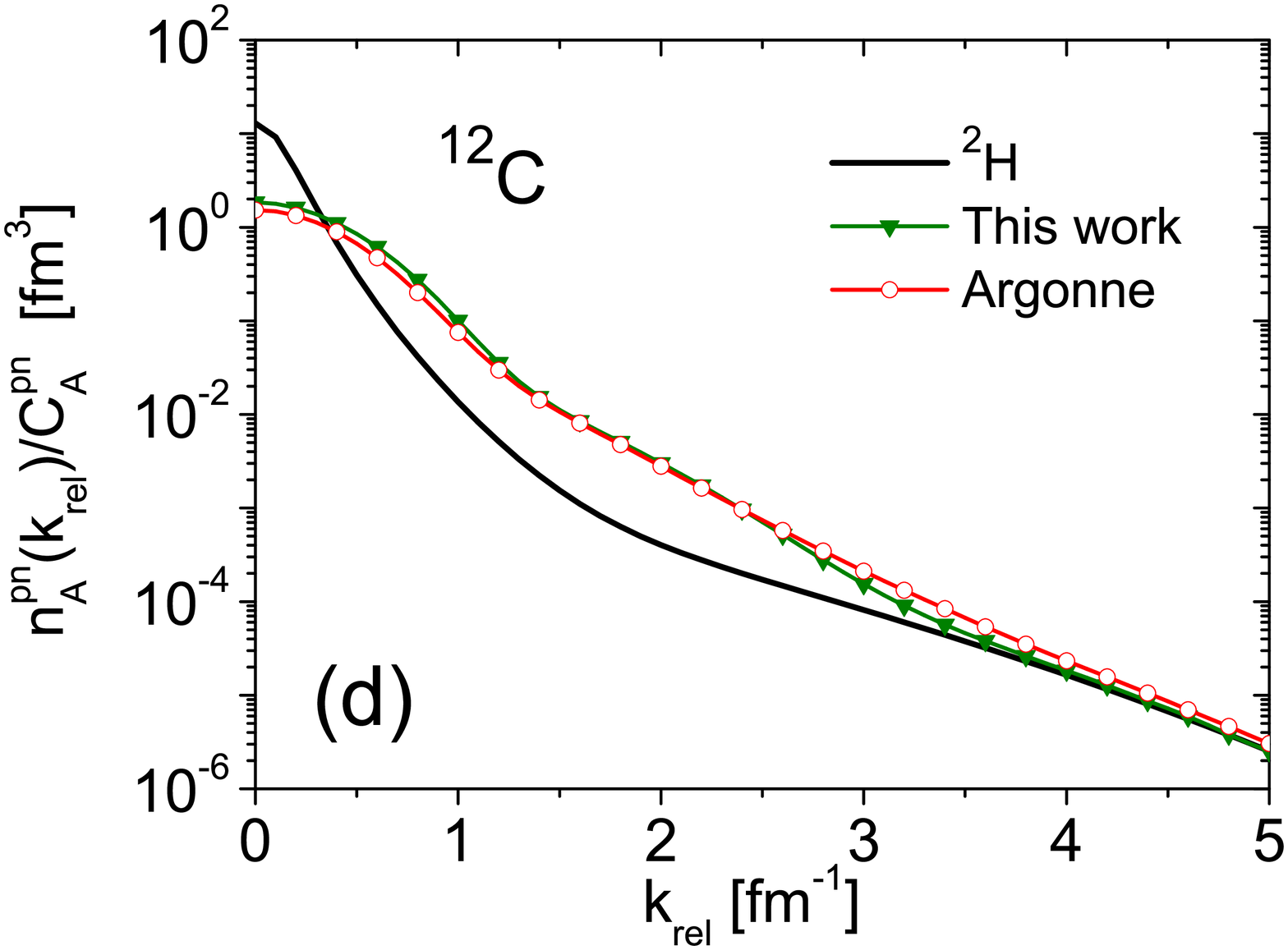}}
  \caption{(Color online) (UPPER PANEL): Comparison of the deuteron momentum distributions with the
    two-nucleon momentum distributions  of  various nuclei obtained in the present
    paper. Fig.(a) demonstrates the validity of the relation
    $n_A^{pn}(k_{rel}, K_{c.m.})/C_A^{pn}n_{c.m.}^{pn}(K_{c.m.}=0)\simeq n_D(k_{rel})$,
    when $K_{c.m.}=0$ and   $k_{rel}\geq  \sim 2\, fm^{-1}$, whereas  Fig.(b) demonstrates
    that when  the $K_{c.m.}$-integrated two-nucleon momentum distributions are considered
    the relation $n_A^{pn}(k_{rel})/C_A^{pn}\simeq n_D(k_{rel})$ is also valid but only at
    $k_{rel}\geq 3.5 \sim 4\, fm^{-1}$. (LOWER PANEL): The quantity
    $n_{A}^{pn}(k_{rel})/C_A^{pn}$ for $^4$He (c) and $^{12}$C corresponding to the
    momentum distributions obtained in  the present paper and in Ref. \cite{Wiringa:2013ala}.
    In both panels the values of  $C_A^{pn}$ are the ones given in Table \ref{Table1}. These results unambiguously
    prove  both $n_A^{pn}({k}_{rel},{K}_{c.m.}=0)$ and $n_A^{pn}(k_{rel})$ do factorize to the
    deuteron momentum distribution but starting at appreciably different values of  $k_{rel}$ in the
    two cases. These results also show that $n_A^{pn}(k_{rel})$ at $k_{rel}\geq 3.5 \sim 4\, fm^{-1}$
    is mainly governed by back-to-back $pn$ pairs.}
  \label{Fig12}
\end{figure*}
 As already pointed out, the in-medium $pn$ momentum distribution is a  relevant quantity  for the study of  in-medium dynamics since
it represents a unique opportunity  to compare the properties of a free bound $pn$ system with the properties of a $pn$ system embedded in
the medium.  The ratio \beqy R_{pn/D}(k_{rel},K_{c.m.}=0)=\frac{n_A^{pn}(k_{rel}, K_{c.m.}=0)} {C_A^{pn} n_{c.m.}^{pn}(K_{c.m.}=0)}
\label{ratio_pn0/D} \eeqy is presented in Fig.\ref{Fig12}(a), whereas  the quantity
\beqy
R_{pn/D}(k_{rel})=\frac{n_A^{pn}(k_{rel})}{C_A^{pn}} \label{ratio_pn/D}
\eeqy
is shown in  Figs.\ref{Fig12}(b), (c) and (d). The scaling of
the Eq. (\ref{ratio_pn0/D}) to the deuteron momentum distributions, starting from $k_{rel}\simeq 2\, fm^{-1}$ is clearly exhibited and it
can also be seen that   scaling  of $n_A^{pn} (k_{rel})$ also takes place ({\it cf} Eq (\ref{ratio_pn/D}), but only at very large values of
$k_{rel}\gtrsim 4\,fm^{-1}$. These results are both obtained with  our momentum distributions and with the VMC ones. By
comparing  Figs. \ref{Fig12}(a) and (b)  it can  be concluded  that the $pn$ momentum distribution in nuclei is governed, at high value of
the relative momentum,  only by the deuteron-like  momentum components, \ie by the two-nucleon  momentum distributions with $K_{c.m.}=0$.
\subsection{Wave function factorization and the relation between the one-nucleon and the
  two-nucleon momentum distributions. The one-nucleon momentum distribution {\it vs} the deuteron
  momentum distribution}
The results presented in Fig. \ref{Fig10} and Fig. \ref{Fig11} represents unquestionable evidence
of the validity of the factorization property, which leads to  the convolution model (CONV) of the
one-nucleon spectral function and  momentum distributions describing both quantities in terms
of a convolution integral of the relative and c.m. momentum distributions of a correlated pair \cite{CiofidegliAtti:1995qe}.  Within the
CONV the  exact relation between the one- and two-nucleon momentum distributions, namely (\textit{e.g.} for protons)
\beqy
&&\hspace{-1cm}n_A^{p}({\bf k}_1)=\frac{1}{A-1}\left( \int n_A^{pn}({\bf k}_1,{\bf k}_2)\,d\,{\bf k}_{2}\nonumber\right.\\
&&\left.\hspace{3cm}+2\int n_A^{pp}({\bf k}_1,{\bf k}_2)\,d\,{\bf k}_{2}\right)
\label{enne1_general}
\eeqy
is represented
 \textit{in the correlation region} at high momenta by the following
convolution integrals (${\bf k}_1+{\bf k}_2 +{\bf k}_3=0$, ${\bf k}_3 = {\bf K}_{A-2}= -{\bf K}_{c.m.}=-({\bf k}_1+{\bf k}_2$))
\beqy
n_A^{p}({\bf k}_1)&&=
\int n_{rel}^{pn}(|{\bf k}_1-\frac{{\bf K}_{c.m.}}{2}|) \, n_{c.m.}^{pn}({\bf
K}_{c.m.})\, d\,{\bf K}_{c.m.}\nonumber\\
&&+2\int n_{rel}^{pp}(|{\bf k}_1-\frac{{\bf K}_{c.m.}}{2}|) \, n_{c.m.}^{pp}({\bf K}_{c.m.})\, d\,{\bf K}_{c.m.}.\nonumber\\
\label{enne1}
\eeqy
\begin{figure*}[!htp] 
  \centerline{\includegraphics[width=0.55\textwidth]{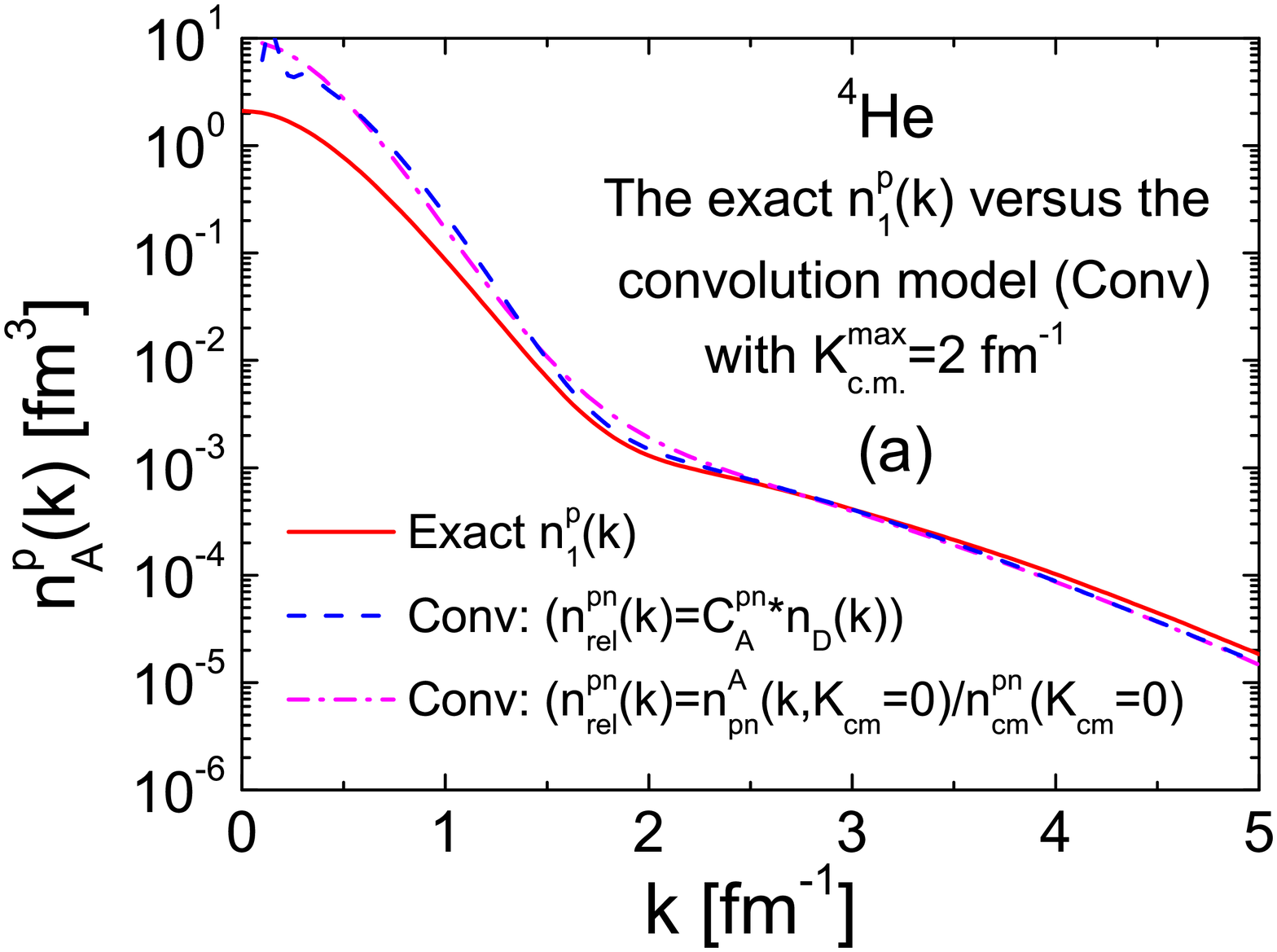}
  \hspace{-1.0cm}
  \includegraphics[width=0.55\textwidth]{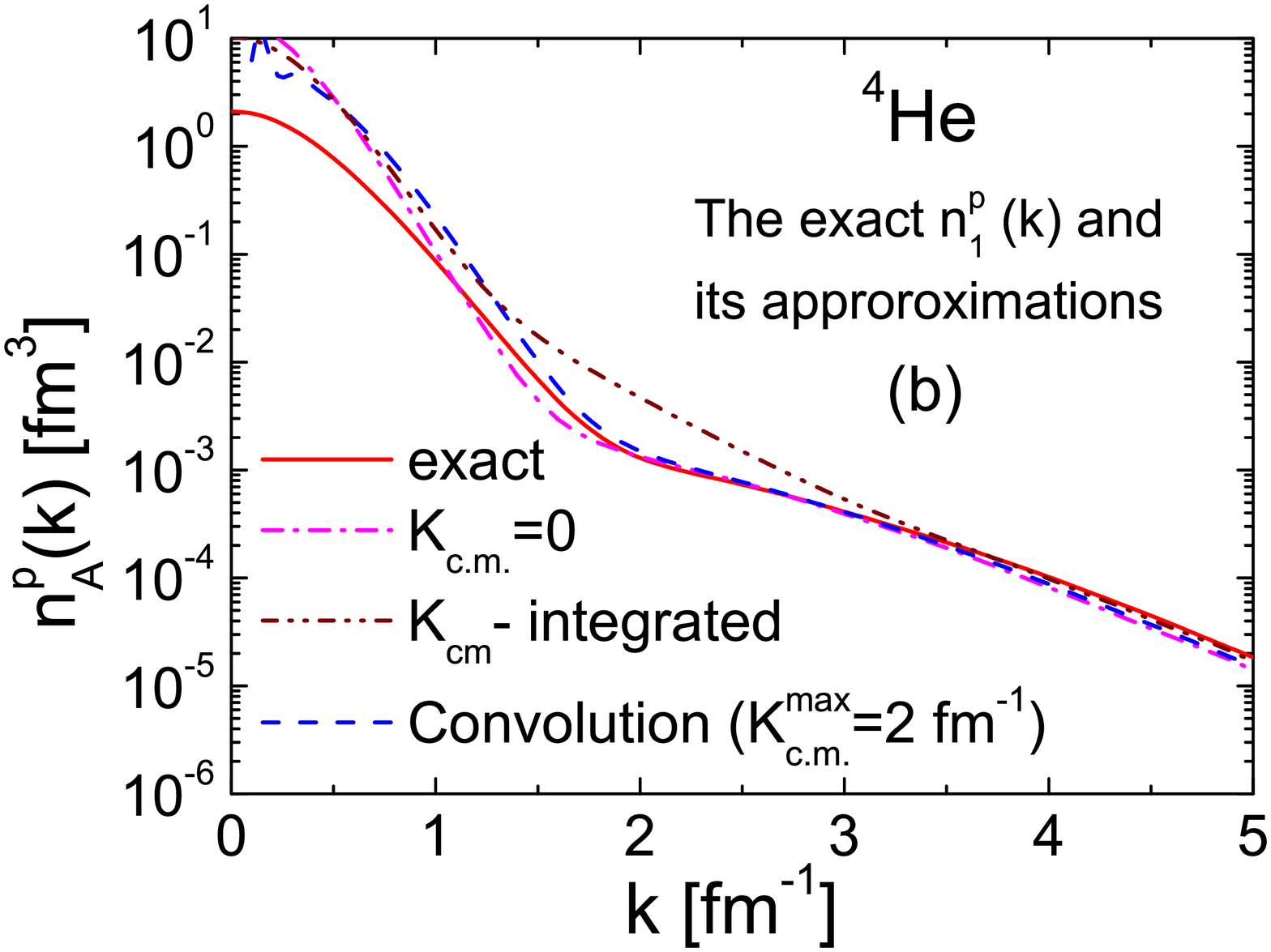}}
  \caption{(Color online) (a) The exact proton momentum distribution
  $n_A^{p}(k)$ ($k_1\equiv k$)
  compared with  the convolution  model,  Eq.(\ref{enne1}) (Conv), calculated with two different
  expressions for $n_{rel}^{pn}$.
  (b) The exact $n_A^p(k)$ compared with: (i) the asymptotic approximation of
  convolution model  (Eq. (\ref{n1=nrel}) and Eq. (\ref{n1n12})) ($K_{c.m.}=0$); (ii)
 Eq. (\ref{n1=nrel})  with  $n_{rel}^{N_1N_2}(k_1=k_{rel})$ replaced by
  the $K_{c.m.}$-integrated relative momentum distributions $n_{{A}}^{N_1N_2}(k_{rel})=
  \int n_{A}^{N_1N_2}({\bf k}_{rel},{\bf K}_{c.m.})\,d\,{\bf K}_{c.m.}$;  (iii) the convolution
  model (Eq.(24)) as in (a).}
  \label{Fig12_bis}
\end{figure*}
  Eq. (\ref{enne1})  establishes a relation between the one-nucleon momentum distribution $n_A^p({\bf
k}_1)$ and the  relative and c.m. momentum distributions of the $N_1N_2$ pair
{\footnote{In actual calculations of Ref.
\cite{CiofidegliAtti:1995qe} the exact  Eq. (\ref{enne1}) has been approximated by using  an effective two-nucleon
  momentum distribution.}.
At large values of ${\bf k}_1$, such that
$\Vec{k}_1 >> \Vec{K}_{c.m.}/2$,  the convolution formula could in principle
be approximated  by
\beqy n_A^p(k_1) \simeq n_{rel}^{pn}(k_{rel}=k_1)+2n_{rel}^{pp}(k_{rel}=k_1) \label{n1=nrel} \eeqy which represents the contribution of
back-to-back nucleons to the one-nucleon momentum distribution; Eq. (\ref{n1=nrel})  can also be expressed in the following equivalent form
\beqy
&&\hspace{-1cm}n_A^{p}(k_1)=\frac{n_A^{pn}({k}_{rel}=k_1, K_{c.m.}=0)}{n_{\text{c.m.}}^{pn}(K_{\text{c.m.}}=0)}\nonumber\\
&&\hspace{2cm}+2\frac{n_A^{pp}({k}_{rel}=k_1,
  K_{c.m.}=0)}{n_{\text{c.m.}}^{pp}(K_{\text{c.m.}}=0)}
\label{n1n12}
\eeqy
as well as in the form
\beqy
&&\hspace{-1cm}n_A^{p}(k_1) \simeq C_A^{pn}n_D({k}_{rel}=k_1)\nonumber\\
&&\hspace{2cm}+2\frac{n_A^{pp}({k}_{rel}=k_1, K_{c.m.}=0)} {n_{\text{c.m.}}^{pp}(K_{\text{c.m.}})}\,,
\label{enne12}
\eeqy
which establishes a clear-cut relation between the one-nucleon momentum distribution and the momentum distribution
of the deuteron in case of pairs of nucleons with back-to-back ($K_{c.m.}=0$) momenta
\footnote{Note that in the following, we use $n_{c.m.}^{pp}(K_{c.m.})=n_{c.m.}^{pn}(K_{c.m.})$.}. Starting from a factorized wave
function, a  relation similar to Eq. (\ref{n1=nrel}) has been obtained in Ref. \cite{Weiss:2015mba},
where however, instead of  the relative momentum distribution
$n_{rel}^{N_1N_2}(k_{rel})=n_{A}^{N_1N_2}(k_{rel}=k_1,K_{cm}=0)/n_{c.m.}^{N_1N_2}(K_{c.m.}=0)$, the $K_{cm}-integrated$
relative momentum distributions (Eq.(\ref{n2_rel})) has been used. We will show that, as expected
from Fig. \ref{Fig12}(a) and (b), the relation between the one- and two-body momentum distribution
will be numerically different. Let us first of all analyze the validity of the convolution model.
In Fig. \ref{Fig12_bis} a detailed analysis of the
model is presented  for  the $^4$He nucleus.  The following  features {\it in the region of factorization dominated by SRCs} ($k\gtrsim 2\,
fm^{-1}$), are worth being stressed: (i) the exact momentum distribution $n_A^p$ is correctly approximated  by the convolution formula (Eq.
(\ref{enne1})) and, particularly, by its asymptotic behavior (Eq. (\ref{n1n12})) including its
 deuteron-like character for the $pn$ distribution \ie for back-to-back SRCd nucleon pairs;
 (ii) the exact calculation, the calculation with the
convolution formula, using there either $C_A^{pn}n_D(k_{rel})$ or $n_A^{pn}({k}_{rel}, K_{c.m.}=0)/n_{\text{c.m.}}^{pn}(K_{\text{c.m.}}=0)$
for the relative motion,
 yield very similar
results starting from $k_{rel} \gtrsim 2 \, fm^{-1}$, whereas Eq. (\ref{n1n12}) with the ${\bf K}_{c.m.}-integrated$ relative momentum
distribution reproduce $n_A^p(k)$ only when $k_{rel} \gtrsim 3.5-4 \, fm^{-1}$.
\begin{figure*}[!htp] 
  \centerline{\includegraphics[width=0.55\textwidth]{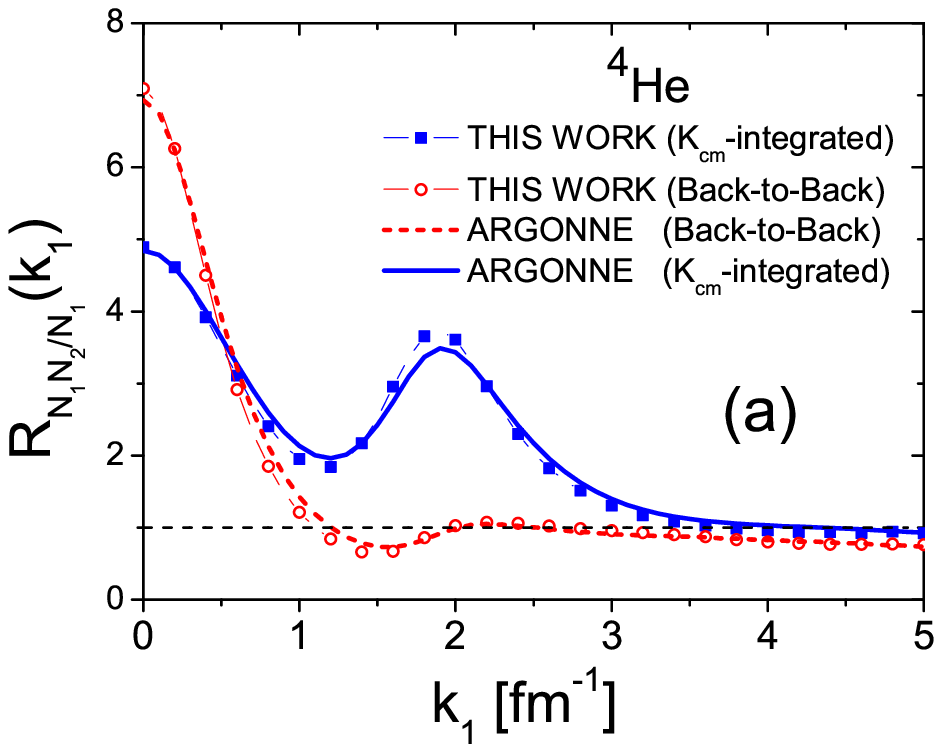}\hspace{-1.0cm}
    \includegraphics[width=0.55\textwidth]{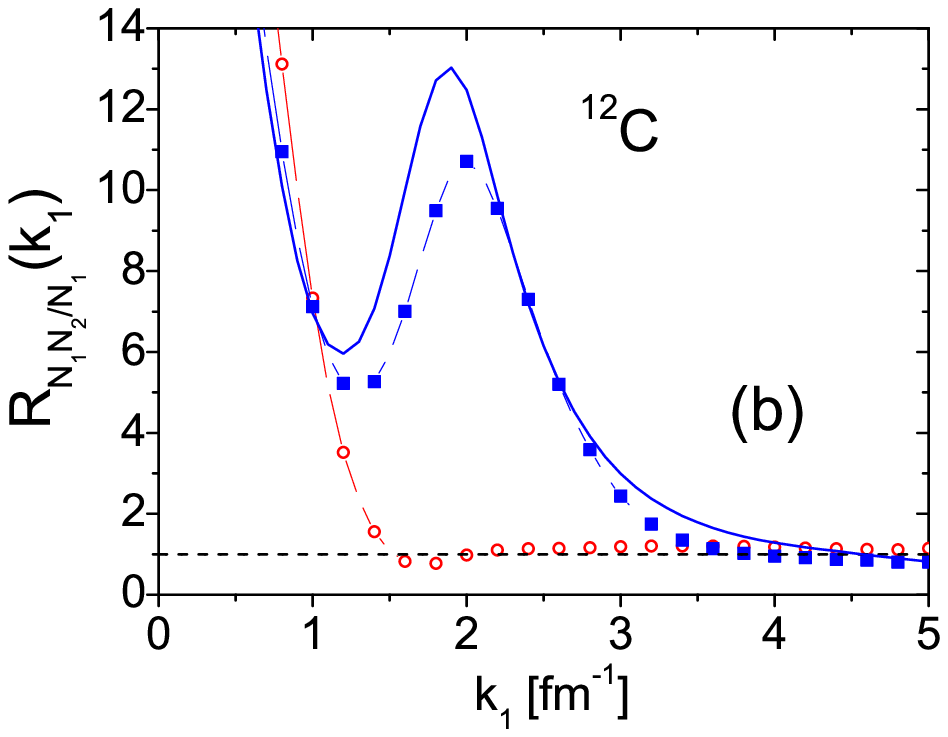}}
  \caption{(Color online) The ratio of the full two-nucleon momentum distribution
    $n_{pn}+2n_{pp}$
    to the one-nucleon momentum distribution $n_A^p$ for $^4$He (a)
    and $^{12}$C (b) calculated using in the numerator the relative two-nucleon  distributions
    $n_A^{pn}(k_{rel}=k_1, K_{c.m.}=0)/n_{c.m.}(K_{c.m.}=0)$ (Eq. (\ref{ratio_enne12}),
   (open dots),
    and (ii)  the ${\bf K}_{c.m.}$-integrated two-nucleon momentum distributions
    $n_{A}^{pn}(k_{rel})$
    (Eq.(\ref{ratio_barnea}), full squares); in both cases the numerator is the one-nucleon momentum
    distribution. The solid line denotes the results obtained with the Argonne VMC
    wave function \cite{Wiringa:2013ala}.}
  \label{Fig13}
\end{figure*}
In order to further demonstrate the relationships of the
one- and two-nucleon momentum distributions we show in Fig. \ref{Fig13} the ratios
\beqy
&&\hspace{-1cm}R_{N_1N_2/N_1}^{BB}(k_1)=
\frac{1}{n_A^p(k)}\Bigl[
  \frac{n_{A}^{pn}(k_{rel},{K}_{\text{c.m.}}=0)}
       {n_{\text{c.m.}}^{pn}(K_{\text{c.m.}=0)}}\nonumber\\
       &&\hspace{2cm}+2\frac{n_{A}^{pp}(k_{rel},{K}_{\text{c.m.}}=0)}
       {n_{\text{c.m.}}^{pp}(K_{\text{c.m.}=0)}} \Bigr]
\label{ratio_enne12}
\eeqy
and \beqy {R}_{N_1N_2/N_1}^{int}(k_1)=\frac{n_{A}^{pn}({k}_{rel}) +2n_{A}^{pp}({k}_{rel})}{n_A^p(k_1)}, \label{ratio_barnea} \eeqy where in
both quantities $n_A^p$ is the exact proton momentum distribution and the numerators differ in that in Eq. (\ref{ratio_enne12})
back-to-back nucleon distributions are considered, unlike the case of Eq.(\ref{ratio_barnea}) where  the ${\bf K}_{c.m.}-integrated$
relative momentum distributions are adopted. The regions of validity of the two cases, both corresponding to $k_1 \simeq k_{rel}$, \ie
$K_{c.m.} \simeq 0$ are determined by a constant value of the ratios. As expected from the results presented in Figs. \ref{Fig12} and
\ref{Fig12_bis}, Eq. (\ref{ratio_enne12}) is unity in a wider range of momenta. The results presented in Fig. \ref{Fig13} provide  further
evidence of the validity of both the factorization property and the convolution model, and tells us that when the ratios equals  one, the
one-nucleon momentum distribution is dominated by back-to-back configurations with ${\bf k}_{1}=-{\bf k}_{2}={\bf k}_{rel}$,  ${\bf
K}_{c.m.}=0$. Concerning the relationship of the one-nucleon momentum distributions and the momentum distributions of the deuteron, as
already illustrated, this  is given by Eq. (\ref{enne12}). However, by plotting the ratio of the one-nucleon momentum distribution
to the  momentum distributions of the deuteron $R_{A/D}(k_1)=\frac{n_A^p(k_1)}{n_D(k_1)}$ the relationships between the two quantities can
be  exhibited in more detail, as quantitatively illustrated in Ref.\cite{Alvioli:2012qa}. There it has been shown that $N_{A/D}(k_1)$ never
becomes constant, which means  that $n_{1}^{A}(k_{1})$ is not linearly proportional to $n_D(k_1)$; this is mostly due to the contribution
of the $pp$ distribution, which increases with increasing momentum $k_1$, and to the c.m. motion of a $pn$ pair in the nucleus  and only if
$pp$ contributions are disregarded and only back-to-back $pn$ pairs are considered, one indeed obtains that in the region $k_1 \gtrsim
2\,fm^{-1}$,
 $n_A^{p}(k_1) \simeq C_A^{pn}n_D(k_1)$.
 The relation between
the  nucleon momentum distribution of nucleus $A$  and the deuteron momentum distribution, has been and is still being
 used in the treatment of SRCs. Still now the proportionality
of the nuclear momentum distribution to the momentum distribution of the deuteron
 is sometimes assumed, which is equivalent to the statement that
the high momentum content of the nucleus is fully determined by the two-nucleon state $(ST)=(10)$. In the past realistic  calculations of
the nuclear momentum distributions at high momenta could not  be performed with sufficient accuracy and the similarity of the deuteron and
the nuclear momentum distribution has been simply assumed, e.g. in the
early VMC calculations \cite{Schiavilla:1985gb} or  in the
development of workable models of the spectral function for complex nuclei \cite{CiofidegliAtti:1995qe}. Recent advanced calculations of
the one- and two-body momentum distributions \cite{Alvioli:2007zz,Alvioli:2011aa,Alvioli:2013qyz,Wiringa:2013ala,Alvioli:2012qa}, including
the results of the present paper,  show that also  states different from the deuteron one, namely the states $(01)$ and $(11)$, do
contribute to the high momentum part of the momentum distributions, demonstrating, in the case of the state $(11)$,  that a considerable
number of two-nucleon  states with odd value of the relative orbital momentum is present in the realistic ground-state wave function of
nuclei.
\subsection{Wave function factorization and the nuclear contacts}
The concept of {\it contact},  introduced by Tan in Ref. \cite{Tan:2008},  to describe the short-range behavior of  two unlike electrons in a
two-component Fermi gas,  has  been recently discussed within the context of SRCs
in nuclei (see e.g. Refs. \cite{Weiss:2015mba,Hen:2014lia}). Although a
detailed discussion of this topic is outside the aim of the present paper
and will be discussed elsewhere, it is
nevertheless useful to stress here that  the
 {\it contacts}: (i) are quantities that measure the probability to find two particles at
short relative distances (\cite{Tan:2008,Weiss:2015mba}) and,  (ii) they are obtained, both in atomic and nuclear systems, by postulating a
factorized wave functions of the form of Eq. (\ref{wf_fact}) \cite{Weiss:2015mba}. For these reasons,  the quantity  $C_A^{pn}$ we have
obtained, measuring  the probability to have SRCd  back-to-back $pn$ pairs, represent  nuclear $pn$ contacts \footnote{It should be
stressed that in case of nuclei  four contacts,
    depending upon the spin-isospin state  of the pair, can be defined; moreover,
     the contacts may be defined to
    depend upon the center-of-mass of the correlated pair, namely for a fixed value of the
    c.m. momentum, for  back-to-back
    nucleons,  as well as  for  the ${\bf K}_{c.m.}$-integrated momentum distributions.}.
    \begin{table*}[!htp] 
  \normalsize
  \begin{center}
    {\renewcommand\arraystretch{1.3}
      \begin{tabular}{|c||c|c|c||c|c|c||c|c|c|}
        \hline
        & \multicolumn{3}{c||}{$^2$H}
        & \multicolumn{3}{c||}{$^3$He}
        & \multicolumn{3}{c|}{$^4$He }\\
        \hline & ${N}_{N_1N_2}^{BB}$&${N}_D^{SRC,BB}$&
        $\mathcal{P}_D^{SRC,BB}$&
        ${N}_{N_1N_2}^{BB}$ &  ${N}_{N_1N_2}^{SRC,BB}$ &
        $\mathcal{P}_{N_1N_2}^{SRC,BB}$& ${N}_{N_1N_2}^{BB}$
        &${N}_{N_1N_2}^{SRC,BB}$ & $\mathcal{P}_{N_1N_2}^{SRC,BB}$ \\
        \hline
        \multirow{2}{*}{{ }pn{ }{ }}& 1  &0.036     &3.6 &    6.22 & 0.22  & {3.5}& 2.54& {0.08} & 3.1\\
        & (1)& (0.036) &(3.6)&  {(5.82)} & (0.20)  & {(3.4)} & {(2.05)} & ({0.09})  & {(4.3)} \\\hline
        \multirow{2}{*}{{ }pp{ }{ }}& -& - & - &    2.05 & {0.01} & {0.5} & {0.55}&  {0.005} &{0.9} \\
        & -& - & - &    {(2.10)} & ({0.01}) & {(0.5)} & {(0.42)}  & (0.004) & {(1.0)} \\
        \hline \hline
        \hline
        & \multicolumn{3}{c||}{$^{12}$C }
        & \multicolumn{3}{c||}{$^{16}$O}
        & \multicolumn{3}{c|}{$^{40}$Ca}\\
        \hline & ${N}_{N_1N_2}^{BB}$& ${N}_{N_1N_2}^{SRC,BB}$ &
        $\mathcal{P}_{N_1N_2}^{SRC,BB}$ &
        ${N}_{N_1N_2}^{BB}$&${N}_{N_1N_2}^{SRC,BB}$&
        $\mathcal{P}_{N_1N_2}^{SRC,BB}$
        &${N}_{N_1N_2}^{BB}$ &
        ${N}_{N_1N_2}^{SRC,BB}$& $\mathcal{P}_{N_1N_2}^{SRC,BB}$\\
        \hline
            { }pn{ }{ }& 3.80 & 0.08& 2.1 & 7.32 & {0.11} &{1.5}& 59.07 & 0.24 & 0.4\\
            { }pp{ }{ }& 1.72 & 0.01 & 0.6 & 3.27&  0.01 & {0.4}& 27.61 & 0.02 & {0.1}\\
            \hline \hline
    \end{tabular}}
  \end{center}
  \caption{The  number of
    of back-to-back (BB) proton-neutron ($pn$)
    and proton-proton ($pp$) pairs (Eq.(\ref{EnneBB_Total})) and the integrated momentum distribution of BB SRCd pairs
    (Eq.(\ref{EnneBB_SRC}))
    and the percent probability ${\mathcal P_{N_1N_2}^{SRC,BB}}=100\,
    N_{N_1N_2}^{SRC,BB}/N_{N_1N_2}^{BB}$.
    Microscopic wave functions
    corresponding to the AV18 interaction \cite{Wiringa:1994wb}, for $^2$H and $^3$He
    \cite{Kievsky:1992um,Kievsky:2010zr}, and to the AV8$^{\prime}$ interaction
    \cite{Pudliner:1997ck} for
    $^4$He \cite{Akaishi:1987} and complex nuclei \cite{Alvioli:2005cz}.
    In brackets the values obtained with the VMC momentum distributions  of
    Ref. \cite{Wiringa:2013ala}, which are calculated with AV18+UX interaction.}
  \label{Table2}
\end{table*}
%
\section {On the number   of  high-momentum short-range correlated  nucleon-nucleon pairs in nuclei} \label{Sec:5}
\begin{figure*}[!htp] 
 \centerline{\includegraphics[width=0.50\textwidth]{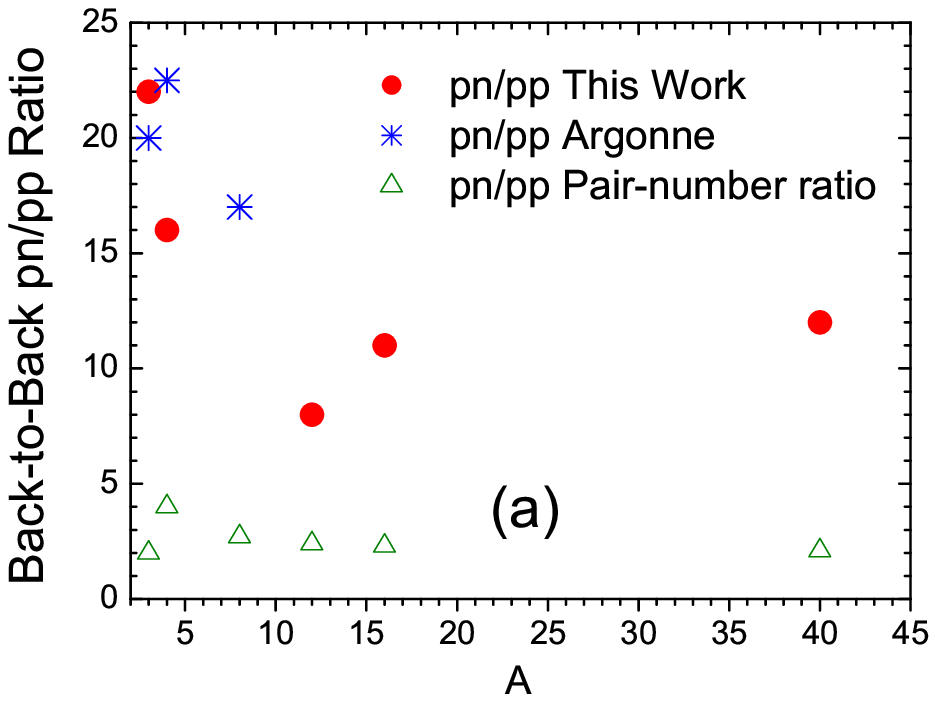}
\includegraphics[width=0.50\textwidth]{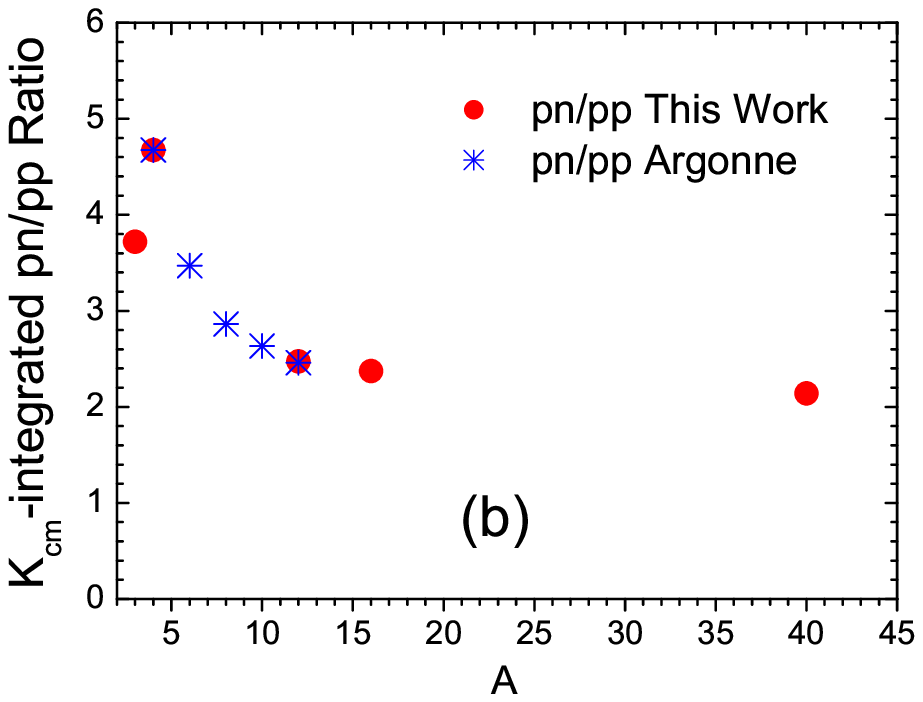}}
 \caption{(Color online) (a): The ratio ${{N}_{pn}^{SRC,BB}/N_{pp}^{SRC,BB}}$ using
   the wave functions of  the present work (\textit{cf.} Table \ref{Table2}) and
   the VMC results of ref \cite{Wiringa:2013ala}. (b):
   the same as in (a) for the $K_{cm}-integrated$
   momentum distributions (\textit{cf.} Table \ref{Table3}).}
 \label{Fig9}
\end{figure*}
Having at disposal the two-nucleon momentum distributions, the absolute values of the number of SRCd pairs, \ie the integral of the
two-nucleon momentum distributions in a given relative and c.m. momentum region, can be calculated and, as in the case of the deuteron, a
proper definition of the probability of SRCs in a nucleus can be given. However in a complex nucleus the two-nucleon momentum distributions
depend upon three variables so that, as pointed out in Ref. \cite{Alvioli:2012qa}, there is a certain degree of ambiguity in providing a
clear-cut definition of the probability of  SRCs in terms of an integral of the two-nucleon momentum distributions.
 In the
case of the deuteron, which is described only in terms of a back-to-back (BB) configuration ($\Vec{k}_1=-\Vec{k}_2 = \Vec{k}$,
$\Vec{k}_{rel} \equiv \Vec{k}$, $\Vec{K}_{\text{c.m.}}=0$), a commonly adopted definition of the  probability  of SRCs is given by the
integral of the   momentum distribution $n_D(k)$ ($k_{rel}\equiv k$) in the interval $1.5 \leq k \leq \infty\,fm^{-1}$, which is the region
dominated by the high momentum components generated by the repulsive core and by the deuteron D-wave produced by the tensor
force.
Therefore in the  deuteron the total number of $pn$ pairs is   $N_D=1$, and   the number of back-to-back (BB) SRCd $pn$ pairs is
\beqy N_{D}^{BB}=4\pi\int_{k^{-}=1.5}^\infty n_D({k})\,k^2\,d\,k \equiv N_{pn}^{BB}
=\simeq 0.036\,, \label{SRC_pairs_D}
\eeqy
\textit{i.e.} only 4\% of the pn pair is  SRCd (such a percentage corresponds to the AV18 interaction).
The extent to which such a
probability will differ in a complex nucleus is a relevant issue, for it can provide information on in-medium effects on short-range $pn$
dynamics. For this reason, a similar definition, \ie the integral of the relative momentum distribution in the range $ k_{rel} \gtrsim
1.5\,\,fm^{-1}$, might  also be introduced in the case of a complex nucleus, keeping however in mind that in a nucleus all possible values
of ${K}_{\text{c.m.}}$ and $\Theta$, as well as all four spin-isospin (ST) values of the pair (mostly $(10)$, $(01)$, and $(11)$),
contribute to the momentum distributions,
 as demonstrated in Refs. \cite{Feldmeier:2011qy,Roth:2010bm},
\cite{Alvioli:2012qa} and \cite{Wiringa:2013ala}. We will consider
the following quantities:
\begin{table*}[!htp] 
  \normalsize
  \begin{center}
    {\renewcommand\arraystretch{1.3}
      \begin{tabular}{|c||c|c|c||c|c|c||c|c|c|c|}
        \hline
        & \multicolumn{3}{c||}{$^2$H}
        & \multicolumn{3}{c||}{$^3$He}
        & \multicolumn{3}{c|}{$^4$He }\\
        \hline & ${N}_{N_1N_2}$&$N_D^{SRC}$&
        $\mathcal{P}_D^{SRC}(\%)$&
        ${N}_{N_1N_2}$ &  $N_{N_1N_2}^{SRC}$ & $\mathcal{P}_{N_1N_2}^{SRC}(\%)$& $N_{N_1N_2}$
        &$N_{N_1N_2}^{SRC}$ & $\mathcal{P}_{N_1N_2}^{SRC}(\%)$ \\
        \hline
        \multirow{2}{*}{{ }pn{ }{ }}&  1  & 0.036   & 3.6 &   2 & {0.093} & {4.7}& 4& {0.243} & {6.1} \\
                        & (1 )& (0.036) &(3.6)& -- & -- & -- & -- & (0.332) & (8.3) \\\hline
        \multirow{2}{*}{{ }pp{ }{ }}& - & -- & -- &  1 & {0.025} & {2.5} & 1&  0.052 & 5.2 \\
                        & -- & -- & -- & -- & -- & -- & -- &  (0.071) & ({7.1}) \\
        \hline \hline
          \hline
          & \multicolumn{3}{c||}{$^{12}$C }
          & \multicolumn{3}{c||}{$^{16}$O}
          & \multicolumn{3}{c|}{$^{40}$Ca}\\
          \hline & ${N}_{N_1N_2}$& $N_{N_1N_2}^{SRC}$ &
          $\mathcal{P}_{N_1N_2}^{SRC}(\%)$ &
          ${N}_{N_1N_2}$&$N_{N_1N_2}^{SRC}$&
          $\mathcal{P}_{N_1N_2}^{SRC}(\%)$
          &${N}_{N_1N_2}$ &
          $N_{N_1N_2}^{SRC}$& $\mathcal{P}_{N_1N_2}^{SRC}(\%)$\\
          \hline
          \multirow{2}{*}{{ }pn{ }{ }}& 36 &3.02 & {8.4}& 64 & 4.75 & 7.4 & 400 &  21.06 & {5.3} \\
               & - &(3.74) &({10.4})& -- & -- & -- & -- & -- & --\\\hline
          \multirow{2}{*}{{ }pp{ }{ }} & 15 & 1.22 & 8.1 & 28 &  2.00 & 7.1& 190 &  9.86 & {5.2}\\
              & - & (1.52) & (10.1) & -- & -- & -- & -- &  -- & --\\
          \hline \hline
      \end{tabular}}
  \end{center}
    \caption{The total number of pairs $N_{N_1N_2}$ (Eq. \ref{TotalNN}), the total number
        of SRCd pairs $N_{N_1N_2}^{SRC}$, Eq.(\ref{EnneSRC}) (in case of
        deuteron Eq.(\ref{SRC_pairs_D})) and their percent probability
        $\mathcal{P}_{N_1N_2}^{SRC,BB}=100 \, N_{N_1N_2}^{SRC}/N_{N_1N_2}$.  Microscopic wave functions corresponding to the
        AV18 interaction \cite{Wiringa:1994wb} for $^2$H and $^3$He \cite{Kievsky:1992um,Kievsky:2010zr}
        and the AV8$^{\prime}$ interaction \cite{Pudliner:1997ck} for $^4$He \cite{Akaishi:1987}
        and complex nuclei \cite{Alvioli:2005cz}. The values in brackets
        correspond to the VMC wave functions of Ref. \cite{Wiringa:2013ala}.}
      \label{Table3}
\end{table*}
%
{\bf 1}.
The total number of  back-to-back
$N_1N_2$ pairs, ${N}_{N_1N_2}^{BB}(K_{c.m.}=0)$,
resulting from the integration of the pair relative momentum and the
total number of the {\it short-range correlated back-to-back}
$N_1N_2$ pairs, ${N}_{N_1N_2}^{SRC,BB}(K_{c.m.}=0,k_{rel}\geq 1.5)$,
that  are given, respectively by
\beqy
&&N_{N_1N_2}^{BB}(K_{c.m.}=0)\nonumber\\
&&=4\,\pi\int_{0}^{\infty}
{n_A^{N_1N_2}({k}_{\text{rel}},{K}_{\text{c.m.}}=0)}\,k_{rel}^2\,
d{k}_{rel} \nonumber\\
&&\equiv N_{N_1N_2}^{BB}
\label{EnneBB_Total}
\eeqy
\beqy
&&N_{N_1N_2}^{SRC,BB}(K_{c.m.}=0,k_{rel}\geq 1.5)\nonumber\\
&&=4\,\pi\int_{1.5}^{\infty}
    {n_A^{N_1N_2}({k}_{\text{rel}},{K}_{\text{c.m.}}=0)}\,k_{rel}^2\,
    d{k}_{rel}\nonumber\\
    &&\equiv N_{N_1N_2}^{SRC,BB}
\label{EnneBB_SRC}
\eeqy
\begin{table*}[!htp] 
  \normalsize
  \begin{center}
    {\renewcommand\arraystretch{1.3}
      \begin{tabular}{|c||c|c|c||c|c|c|}
        \hline \hline
        & \multicolumn{3}{c||}{$^2$H}
        & \multicolumn{3}{c||}{$^3$He}\\
        \hline &$R_{pn} (\%)$& $R_{pp}(\%)$& $R_{pp/pn}$&
        $R_{pn} (\%)$& $R_{pp}(\%)$& $R_{pp/pn}$\\
        \hline
        { }THE{ }{ }& 100&0 &-&   {89.3}  &  {3.16} &{3.54}\\
        { }EXP{ }{ }& -- & -- & -- & -- & -- & --\\
        \hline \hline
                & \multicolumn{3}{c||}{$^4$He}
        & \multicolumn{3}{c||}{$^{12}$C}\\
        \hline &$R_{pn} (\%)$& $R_{pp}(\%)$& $R_{pp/pn}$&
        $R_{pn} (\%)$& $R_{pp}(\%)$& $R_{pp/pn}$\\
        \hline \hline
        { }THE{ }{ }& {93.2}&  {5.43}& 5.83 &{96.2}&{5.01}&5.20\\
        { }EXP{ }{ }&87.0$\pm$14.1 & 3.9 $\pm$ 1.5 & 5.1 $\pm$ 2.6&97.0$\pm$22.1&4.8$\pm$1.0&5.8$\pm$1.5\\
          \hline
          & \multicolumn{3}{c||}{$^{16}$O}
          & \multicolumn{3}{c|}{$^{40}$Ca}\\
          \hline &$R_{pn} (\%)$& $R_{pp}(\%)$& $R_{pp/pn}$&
          $R_{pn} (\%)$& $R_{pp}(\%)$& $R_{pp/pn}$\\
          \hline
          { }THE{ }{ }&{97.9}&5.05&  5.15 & {91.8} &{6.52}& 7.11 \\
          { }EXP{ }{ }& -- & -- & -- & -- & -- & --\\
          \hline \hline
      \end{tabular}}
  \end{center}
    \caption{ The percent ratio of the $pn$ and $pp$ short-range correlated  BB pairs with respect
        to the total number of correlated pairs and
       the percent ratio of $pp$ to $pn$ pairs  (Eq. (\ref{Ratios_Fin})) at $k_{rel}=2.5\,fm^{-1}$, calculated
       using the back-to-back momentum distributions shown in Figs. \ref{Fig1}-\ref{Fig5}.
       Experimental data for $^{12}$C from
       Refs. \cite{Tang:2002ww,Piasetzky:2006ai,Shneor:2007tu,Subedi:2008zz}
       and for $^4$He from Ref. \cite{Korover:2014dma}.}
      \label{Table4}
\end{table*}
In both Eqs. (\ref{EnneBB_Total}) and (\ref{EnneBB_SRC}), whose values are shown in Table \ref{Table2},
the quantity  $n_A^{N_1N_2}({k}_{\text{rel}},{K}_{\text{c.m.}}=0)$ is the one shown in Figs. \ref{Fig1}-\ref{Fig5}.
\footnote{Note that Eqs.(\ref{EnneBB_Total}) and (\ref{EnneBB_SRC}) have the dimension
  of $fm^3$ provided by the c.m. momentum  distribution at $K_{c.m.}=0$. We call them anyway
  {\it number of particles for back-to-back pairs.}}
It can be seen from Table \ref{Table2} that for nuclei with $A>4$ an appreciable
decrease of the percent  probabilities
$\mathcal{P}_{N_1N_2}^{SRC,BB}$} of back-to-back proton-neutron ($pn$) and proton-proton ($pp$)
nucleons does occur  with increasing
values of $A$, which can be explained by the similar values of $n_{A}^{pn}(K_{c.m.}=0)$ for $A \geq 12$
and, at the same time,  the substantial increase of the   value of the number of back-to-back
proton-neutron and proton-proton  nucleons ${N}_{N_1N_2}^{BB}$
(Eq. (\ref{EnneBB_Total}));

\textbf{2}.
The total number of SRCd pairs defined as the integral in the entire region of variation of
$K_{c.m.}$ and in the region of the relative momentum
with $k_{\text{rel}}^- \gtrsim 1.5 \, fm^{-1}$, \textit{i.e.},
\beqy
&&\hspace{-1cm}N_{N_1N_2}^{SRC}(k_{rel}^-=1.5)\nonumber\\
&&=\int_{{1.5}}^{\infty}
d^3\,{k}_{rel} \int_{{0}}^{\infty}\,d^3\,{K}_{c.m.}
n_A^{N_1N_2}({\bf k}_{rel}, {\bf K}_{\text{c.m.}})\nonumber\\
&&=4\pi\int_{{1.5}}^{\infty}\,{k}_{rel}^2\,d\,{k}_{rel}
n_A^{N_1N_2}({k}_{\text{rel}})
\equiv N_{N_1N_2}^{SRC}
\label{EnneSRC}
\eeqy
This quantity is compared with the total number of pairs given by \beqy N_{N_1N_2}= 4\pi\int_{{0}}^{\infty}\,{k}_{rel}^2\,d\,{k}_{rel}
n_A^{N_1N_2}({k}_{\text{rel}}) \label{TotalNN} \eeqy The results are listed in Table \ref{Table3}. The values of
$N_{N_1N_2}^{SRC}(k_{rel}^-=1.5)$  include both two-nucleon SRCs (2NSRCs), as well as many-nucleon SRCs generated by the hard
high-momentum  tail ($K_{c.m.} \gtrsim 1$) of the c.m. distributions. Note, moreover, that the number of SRCd pairs is the largest one in
this case since the entire variation of $K_{cm}$ is taken into account; also worth being stressed is the almost constant value of the
probability for $A \geq 12$ which is due to the same rate of increase of the number
 of correlated pairs and the total numbers of $pN$ pairs
$N_{pN}$. It can be seen from Fig. \ref{Fig12}(b) that in the region $1.5 \lesssim k_{rel} \lesssim 3.5\, fm^{-1}$ the momentum components
with $K_{c.m.} \neq 0$ are important in $n_A^{pn}(k_{rel})$. The main results of Table \ref{Table2} and \ref{Table3} are summarized in Fig.
\ref{Fig9}, whose main features should be stressed as follows:
\begin{enumerate}
\item because of the $pn$ tensor dominance (see Figs. \ref{Fig1}-\ref{Fig5})
 the number of SRCd $pn$ pairs in
  few-nucleon systems and $A<12$ is larger than the number of  $pp$ pairs by about a
   factor
    twenty, whereas in medium-weight iso-scalar nuclei it is  larger by about  a factor
    ten, to be compared with  a factor of
    two  (2Z/(Z-1)), which is predicted by the naive pair-number ratio;
\item when the total,  ${\Vec K}_{c.m.}$-integrated  number of pairs is
  considered, the value
  of the $pn/pp$ ratio strongly decreases to a factor of about two, due to the role played by the
  c.m. high momentum components, as it can  easily be understood by comparing
  Figs. \ref{Fig1}(c)-\ref{Fig2}(c)  with Figs.
   \ref{Fig1}(d)-\ref{Fig2}(d) and Figs. \ref{Fig3}(b)-\ref{Fig5}(b)
    with Figs. \ref{Fig3}(c)-\ref{Fig5}(c).
\end{enumerate}
\begin{figure*}[!htp] 
  \centerline{\includegraphics[width=0.53\textwidth]{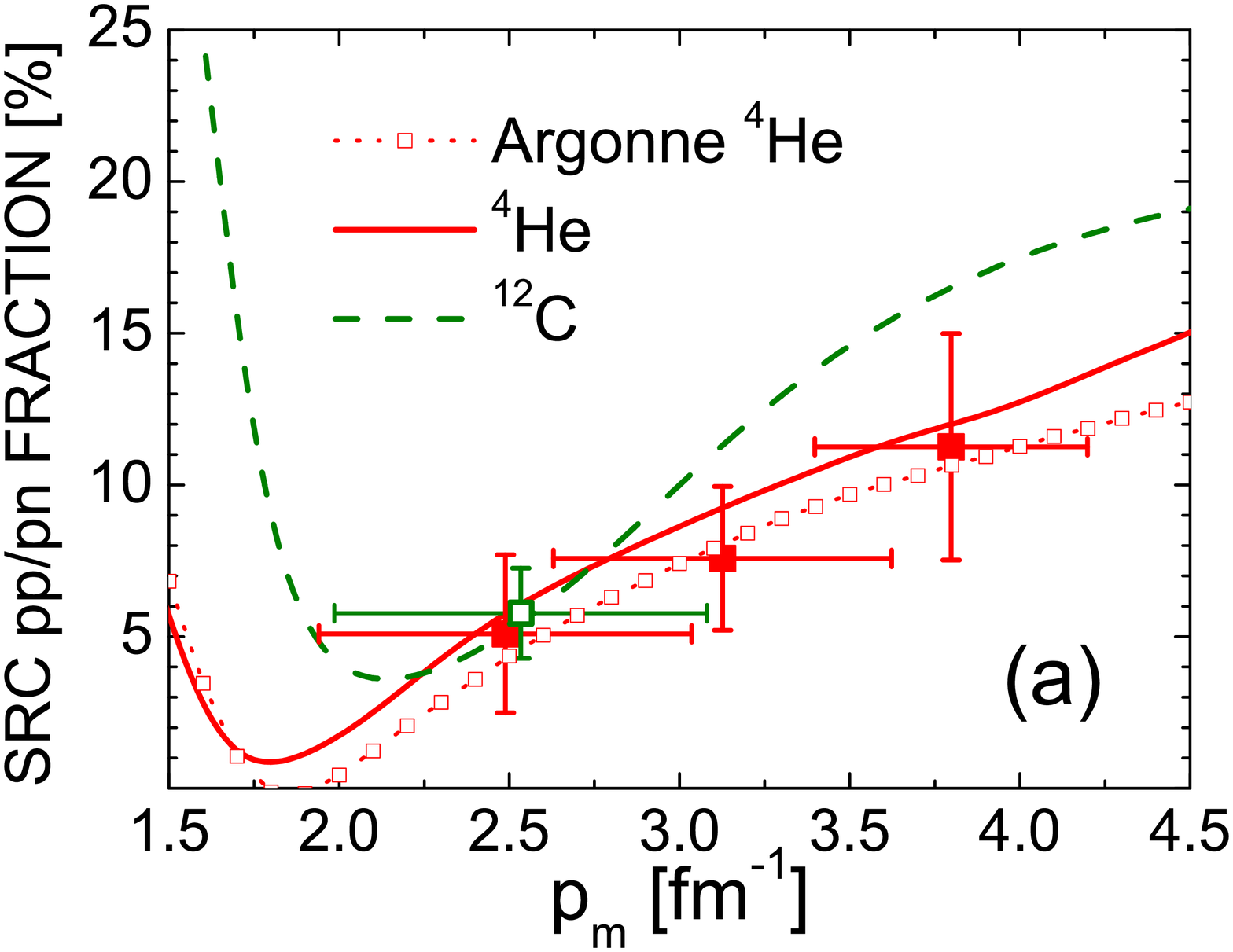}
 \includegraphics[width=0.53\textwidth]{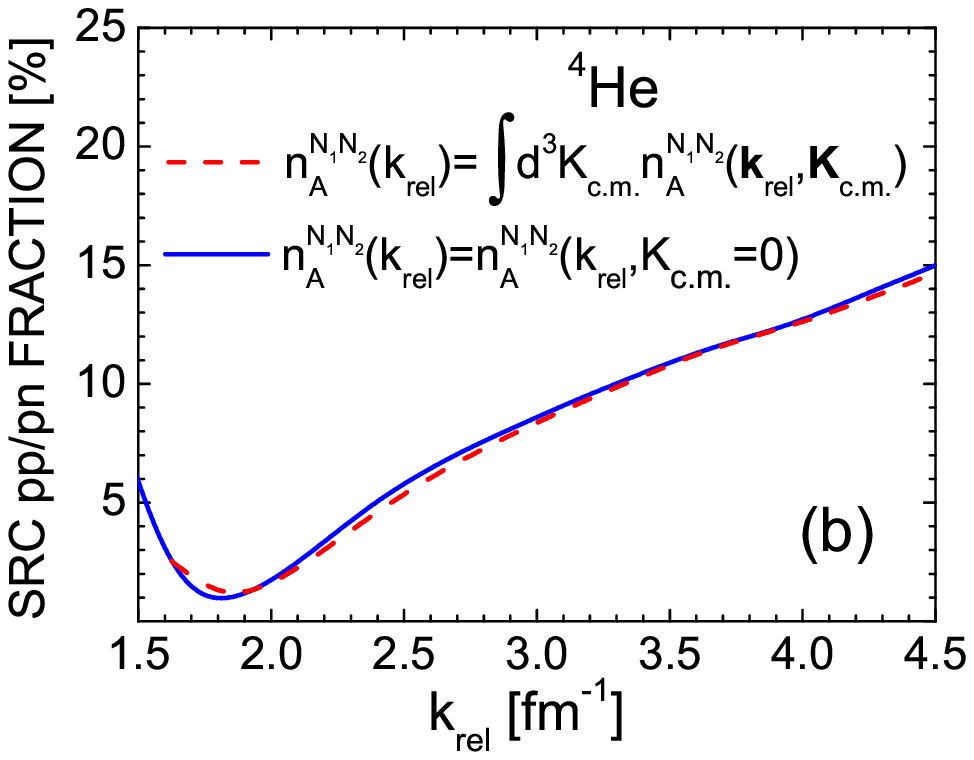}}
  \caption{ (Color online) (a): the  experimental percent  of the $pN$ BB pair fraction $pp/pn$, {\it vs}
    the missing momentum $p_m$, extracted from the processes $^4He(e,e'pn)X$ \cite{Korover:2014dma} and
    $^{12}$C(e,e$^\prime$pp)X \cite{Subedi:2008zz,Shneor:2007tu} compared with
    the quantity $R_{pp/pn}(k_{rel},0)=n_{pn}(k_{rel}, K_{c.m.}=0)/n_{pp}(k_{rel}, K_{c.m.}=0)$
    calculated in the present paper (full line). The open squares show the results obtained with
    the Argonne momentum distributions  \cite{Wiringa:2013ala}. (b): the same ratio
    as in (a) calculated within two different approaches:(i) full line:
    $n_{A=4}^{pn}(k_{rel}, K_{c.m.}=0)/n_{A=4}^{pp}(k_{rel}, K_{c.m.}=0)$; (ii) dashed line:
    $\frac{\int_0^{\infty} n_{pp}({k}_{rel},{K}_{c.m.})
    K_{c.m.}^2\,d\,{K}_{c.m.}}{\int_0^{\infty} n_{pn}({k}_{rel},{K}_{c.m.})
    K_{c.m.}^2\,d\,{K}_{c.m.}}$.
  }
\label{Fig15}
\end{figure*}
\begin{figure*}[!htp] 
  \begin{center}
    \centerline{\includegraphics[width=0.55\textwidth]{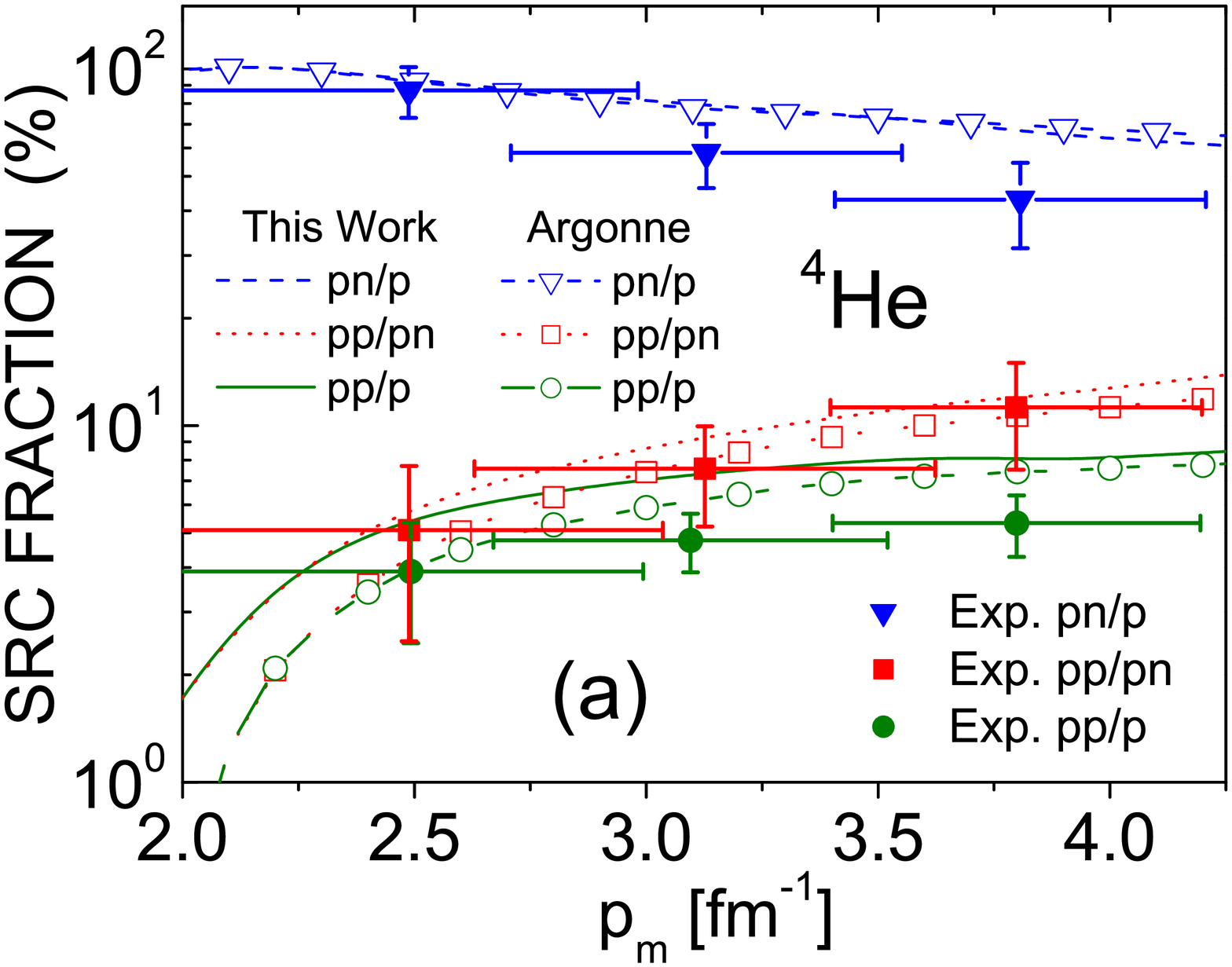}\hspace{-1.0cm}
      \includegraphics[width=0.55\textwidth]{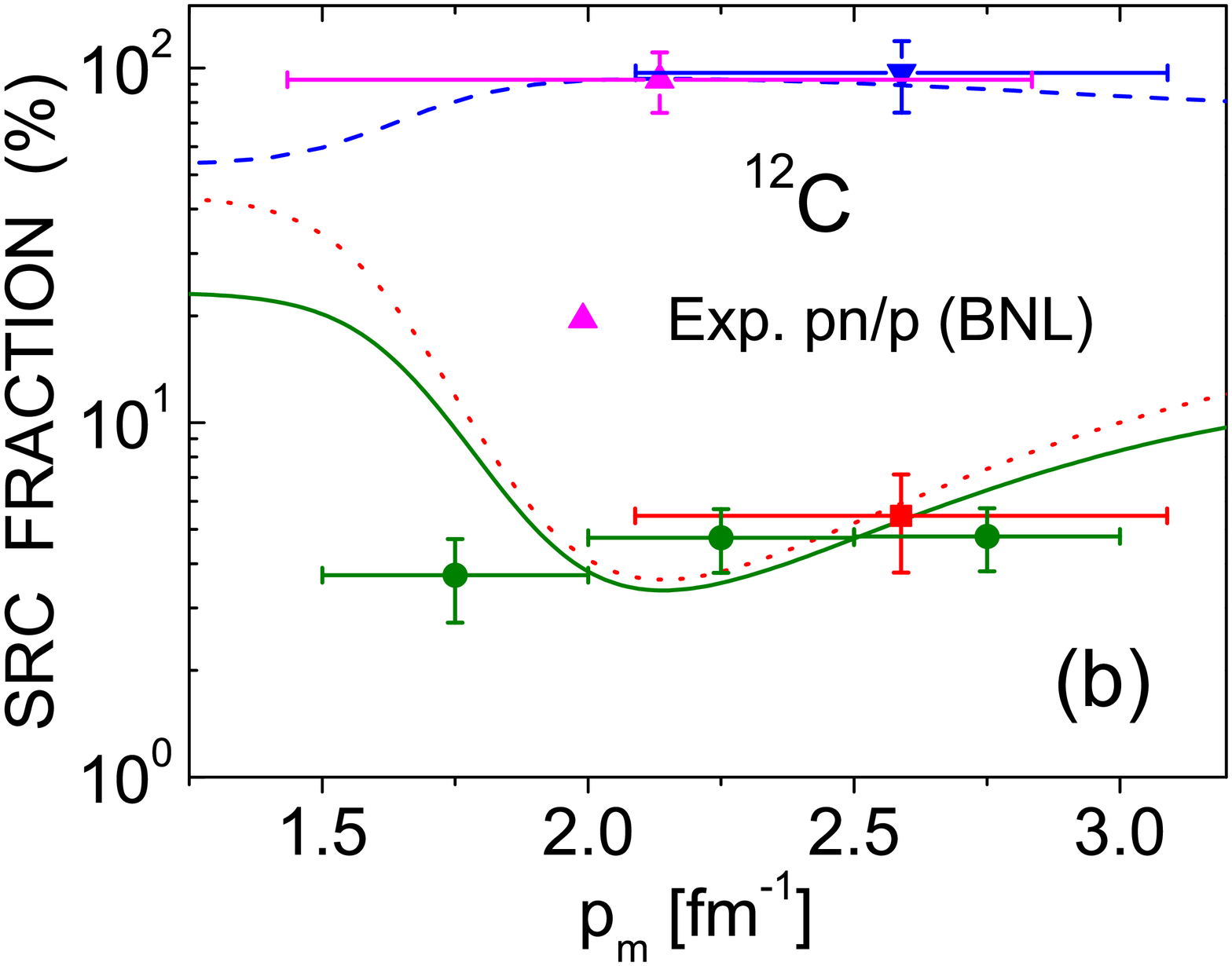}}
    \vskip -0.5cm
    \caption{(Color online) The experimental percent of SRC  fractions in $^4$He (a) and $^{12}$C (b)
      compared with theoretical ratios of momentum distributions within the assumption
      ${\bf p}_{m}\simeq {\bf k}_{rel}$ and $K_{c.m.}=0$. Momentum distributions from
      the present work and from Argonne VMC calculation \cite{Wiringa:2013ala}. All
      experimental data are from Jlab (\cite{Piasetzky:2006ai}-\cite{Hen:2014nza}),
      except the one represented by  the magenta point for $^{12}$C that was obtained
      BNL \cite{Tang:2002ww}. In (b) the three theoretical curves have been obtained in the present
      work and correspond to $pp/p$ (full), $pn/p$ (dashed) and $pp/pn$ (dot-dashed), respectively.}
    \label{Fig16}
  \end{center}
\end{figure*}
\section{Short-range correlations: theoretical predictions  {\it vs} experimental data}
\label{Sec:6}
Experimental investigation of SRCs is a complicated task mainly due to the small value of the involved cross sections and the effects of
FSI that makes it difficult to reconstruct the initial correlated state. Nonetheless, experimental progress  has been recently achieved
thanks to the use of intense lepton beams and the development of advanced detector techniques.  Nowadays it became
possible to investigate quasi-elastic $A(e,e'N_1)X$ and $A(e,e'N_1N_2)X$ processes at high value of $Q^2$ and  Bjorken scaling variable
$x_B >1$, a region where: (i) the contribution from non-nucleonic degrees of freedom is suppressed, (ii) the effects from initial-state
SRCs are emphasized (see Ref.\cite{Frankfurt:2008zv}-\cite{Atti:2015eda}),  and (iii) the theoretical treatment of FSI has reached
high degree of sophistication \cite{Frankfurt:1996xx,Mardor:1992sb,Ryckebusch:2003fc,Cosyn:2007er}. Several SRC properties that
have been experimentally
investigated deserve a  comparison with theoretical calculations which is presented here-below.

\subsection{The percent ratios of different  kinds of  $N_1N_2$ pairs in $^4$He and $^{12}$C and their missing momentum dependence}
\label{Sub:5.1}
SRCs in $^4$He and $^{12}C$   have been recently investigated
\cite{Tang:2002ww,Piasetzky:2006ai,Shneor:2007tu,Subedi:2008zz,Korover:2014dma,Hen:2014nza}
within the following  kinematical region \footnote{The same notations as in Ref. \cite{Atti:2015eda} are
  adopted here}: the squared four-momentum transfer  $Q^2 \simeq 2\,(GeV/c)^2$, the bjorken scaling variable
$x_{Bj}=1.2$ and the three-momentum transfer  $|{\bf q}|\equiv q \simeq 1.6 \,GeV/c$. Information on the
short-range momentum distribution of correlated pairs has been obtained by the following procedure: triple
coincidence processes $^{12}C(p,p^{\prime}pN)X$ and $^{12}C(e,e^{\prime}pN)X$ have been performed by
detecting, in coincidence with the struck,  leading  protons with high momentum ${\bf p}$, protons and neutrons
moving with recoil momentum   ${\bf p}_{rec}={\bf q}-{\bf p}$  along  a direction that, within the plane wave
impulse approximation (PWIA),   would coincide with the direction opposite to the momentum  that the struck nucleon
had before interaction with the projectile. Specifically, within the  PWIA, if before interaction the struck proton
had  a momentum ${\bf k}_1$, the leading proton would have a  momentum ${\bf p}= {\bf k}_1 + {\bf q}$ and
the known {\it missing momentum} would be ${\bf p}_{m} = {\bf q} - {\bf p}=-{\bf k}_1$. Therefore, if  the
struck proton "1" was partner of a correlated nucleon "2"  with momentum ${\bf k}_2 \simeq -{\bf k}_1$, in
coincidence with the leading proton  a recoiling nucleon "2" with momentum ${\bf p}_{rec}={\bf p}_{m}=-{\bf k}_1 \simeq {\bf k}_2$ should
be observed along the direction opposite to ${\bf p}_{m}$. In Refs.
\cite{Tang:2002ww,Piasetzky:2006ai,Shneor:2007tu,Subedi:2008zz,Korover:2014dma,Hen:2014nza} the processes  $A(p,p^{\prime}p)X$,
$A(e,e^{\prime}p)X$\textcolor{blue}{,} $A(p,p^{\prime}pp)X$, $A(e,e^{\prime}pn)X$ and $A(e,e^{\prime}pp)X$ have been investigated  by
detecting mainly back-to-back $pp$ and $pn$ nucleons  in the range $ 1.5 \lesssim p_{m} \lesssim 3\,\,fm^{-1}$ in $^{12}$C, and $1.5
\lesssim p_{m} \lesssim 4\,\,fm^{-1}$  in $^4$He. Within such a kinematic set-up, the percent ratios of the cross sections pertaining to
$pn$ and $pp$ pairs have been extracted. Using  the two-nucleon relative momentum distributions shown in Figs. \ref{Fig1}-\ref{Fig5}
corresponding to BB nucleons ($K_{c.m.}=0$), we have  calculated the following quantities:
\beqy
R_{pn}(k_{rel})&=&\frac{n_A^{pn}}{n_A^p} \equiv \frac{pn}{p};\nonumber\\
R_{pp}(k_{rel})&=&\frac{n_A^{pp}}{n_A^p}\equiv \frac{pp}{p};\nonumber\\
R_{pp/pn}(k_{rel})&=&\frac{n_A^{pp}}{n_A^{pn}} \equiv\frac{pp}{pn}\label{Ratios_Fin}
\eeqy
where $n_A^{pn}\equiv n_A^{pn}({k}_{rel},{K}_{c.m.}=0)/n_{c.m.}^{pn}(K_{c.m.}=0)$ and $n_A^{pp}\equiv
n_A^{pp}({k}_{rel},{K}_{c.m.}=0)/n_{c.m.}^{pp}(K_{c.m.}=0)$. Here $n_A^{N_1N_2}$ is related to the
process $A(e,e'N_1N_2)X$ and $n_A^{p}$ to the process $A(e,e'p)X$, which includes the contributions from
$pn$ and $pp$ SRCs according to Eq. (\ref{enne1_general}), therefore the ratios $pn/p$ and $pp/p$
represent essentially the percent ratios of the SRCd $pp$ and $pn$ pairs with respect  to the total number
of SRCd pairs. The quantities in Eq. (\ref{Ratios_Fin}) have been compared with the experimental data by
assuming  that $p_{m} \simeq k_{rel}$, a procedure that implies the validity of the PWIA, or, at least, the
cancellation of the FSI  in the ratios.
The comparison is presented in  Table \ref{Table4} and in Figs. \ref{Fig15} and \ref{Fig16}.
A general agreement between theoretical and experimental percent ratios appears to hold. Since the experiments have been performed in a
momentum region where factorization of the wave functions is at work, the effects of the c.m. motion largely cancel out in the ratios. As
for the effects of the FSI the experimental kinematics set-up is compatible with the assumption of FSI  effects confined within the
correlated pair, leading also in this case to some kind of cancelation in the ratio (see e.g.
\cite{Frankfurt:1996xx,Mardor:1992sb,Arrington:2011xs,Atti:2015eda,Ryckebusch:2003fc,Cosyn:2007er}).
Concerning the  results
presented in these Figures   the following comments are in order:
\begin{enumerate}
\item our results for $^4$He do not practically differs from the ones obtained with the Argonne distributions;
\item in $^4$He the increase with   $p_{m}=|{\bf p}_{m}|$ of the $pp/pn$ ratio can be explained with the
increasing role of the repulsive NN interaction with respect to the
tensor one ({\it cf } Fig. \ref{Fig2}(c));
However, in spite of this satisfactory agreement, an advanced theoretical approach including  FSI is desirable; preliminary results from
Ref. \cite{Ryckebusch:2003fc,Cosyn:2007er}, quoted in \cite{Korover:2014dma} seem to correct the PWIA into the right direction;
\item the results presented  in Fig. \ref{Fig15}(b) show  that  the ratio calculated at $K_{c.m.}=0$ or integrated  by averaging
  over all direction of ${\bf K}_{c.m.}$ practically do not differ, which is another manifestation of factorization since the $pp$
  and $pn$ c.m. momentum distributions are essentially the same.
\end{enumerate}
\begin{figure*}[!htp] 
\centerline{\includegraphics[width=1.0\textwidth]{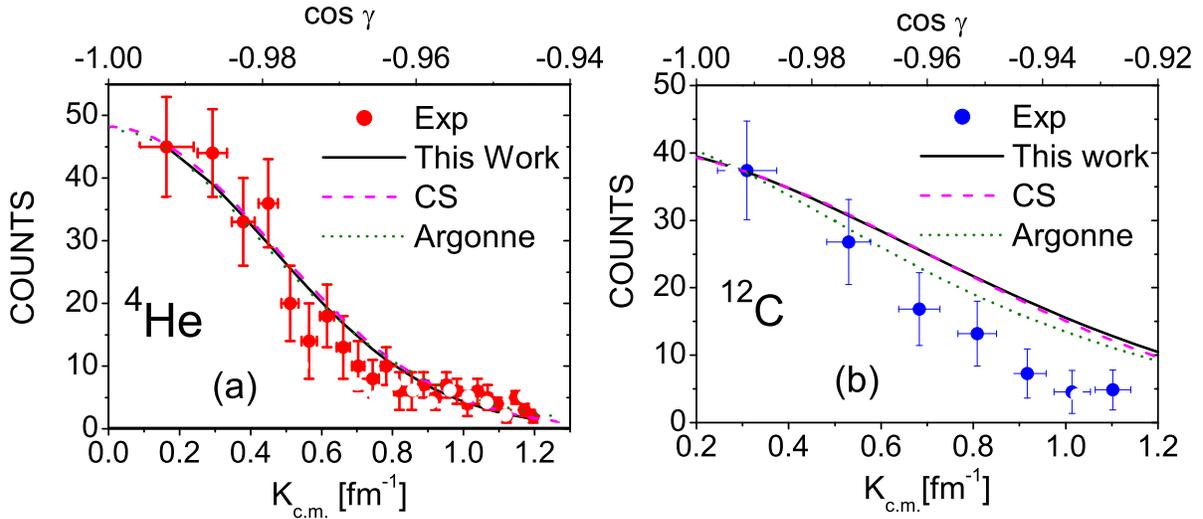}}
  \caption{
   (Color online)   The   c.m. momentum distribution
    of a $pn$ pair in $^4$He (a) and a $pp$ pair in $^{12}$C (b) extracted
    in Refs.  \cite{Korover:2014dma} and  \cite{Shneor:2007tu} from the processes
    $^4He$(e,e$^\prime$pn)X and $^{12}$C(e,e$^\prime$pp)X. $\gamma$ is the angle between
    ${\bf p}_{m}$ and  ${\bf p}_{rec}$, which in PWIA is the angle between   ${\bf k}_{1}$
    and  ${\bf k}_{2}$. The values of ${\bf K}_{c.m.}$ have  been obtained assuming
    ${\bf k}_{2}=-{\bf k}_{1}$. The theoretical curves correspond to the
    momentum distributions of Ref. \cite{Wiringa:2013ala} (Argonne)   and
    \cite{CiofidegliAtti:1995qe} (CS). The experimental data are  given in arbitrary units and the
    theoretical calculations were normalized at the lowest available experimental point. Note that
    the discrepancy between the experimental data  and the  theoretical calculations in the case of
    $^{12}$C is not a real one, since  the latter, unlike the former,  do not take into account the
    finite acceptance and resolution of the detectors \cite{Hen_PC}. Indeed, when these are taken
    into account, the data can be explained by a Gaussian distribution
    $n_{c.m.}(K_{c.m.})=(\alpha/\pi)^{1.5}\,exp(-\alpha K_{c.m.}^2)$ with $\alpha_{exp}= 0.97 \pm 0.19 \,fm^{2}$
    \cite{Shneor:2007tu} in agreement with the three curves in Fig. \ref{Fig17}(b).}
  \label{Fig17}
\end{figure*}
\subsection{The c.m.  momentum distribution of correlated pairs in $^4$He and $^{12}$C} \label{Sub:5.2}
The c.m.  momentum distributions of a correlated ${pn}$ pair relative to the
spectator nucleus $A-2$   in  $^4$He and $pp$ pair in  $^{12}$C  has been
determined  in Refs. \cite{Shneor:2007tu} and \cite{Korover:2014dma} by analyzing the distribution
 of events in the process $A(e,e'pN)X$
as a function of the cosine of the opening angle $\gamma$ between ${\bf p}_{m}$ and
 ${\bf p}_{rec}$ which, in PWIA, is the angle between
${\bf k}_{1}$ and ${\bf k}_{2}$. The results of the analysis of the experimental data, corrected for the detector acceptance, are shown in
Fig.\ref{Fig17} where the theoretical momentum distributions are also shown.  It turns
out  \cite{Hen_PC} that, once the theoretical curves are corrected taking into account
the finite acceptance of the detectors they nicely agree with the experimental
momentum distributions.
\section{Summary and conclusions}\label{Sec:7}
In this paper we have investigated  in-medium short-range nucleon-nucleon dynamics  by calculating various kinds of
  two-nucleon momentum distribution in few-nucleon system and selected isoscalar nuclei with $A \leq 40$. To this end calculations have been
  performed within a parameter-free many-body approach which, even if not fully {\it ab initio}, turned out to be capable to treat high
  momentum components in nuclei with  $A \geq 12$, for which advanced VMC approaches with
  bare strongly repulsive  local interactions, are unfortunately
  not yet feasible. The method,    based upon  a linked cluster expansion of one- and two-nucleon, diagonal and non-diagonal, density
  matrices,
  has been previously used to calculate the ground-state energy \cite{Alvioli:2005cz} and  the
  momentum distributions \cite{Alvioli:2007zz,Alvioli:2012qa}. In this paper
  we have performed  a detailed analysis of  the two-nucleon  momentum
  distributions $n_A^{N_1N_2}(k_{rel}, K_{c.m.},{\Theta})$ at various values
   of ${k}_{rel}$, ${K}_{c.m.}$ and ${\Theta}$, as well as of   the
  two-nucleon relative, $n_A^{N_1N_2}(k_{rel})$, and center-of-mass,
  $n_A^{{N_1N_2}}(k_{c.m.})$,
  momentum distributions of proton-neutron   and proton-proton pairs.
The results of our calculations show that a fundamental  property of the  nuclear wave
function at short inter-nucleon separations turns out to be  its factorization into the relative and the c.m. coordinates, a property which
has been  previously theoretically illustrated in the case of nuclear matter \cite{Baldo:1900zz} and
few-nucleon systems \cite{CiofidegliAtti:2010xv}. Such a property is a very relevant one, for it
fully governs the high momentum behavior of two-nucleon momentum distributions generated by short-range correlations. In particular, the following
properties of in-medium two-nucleon dynamics, resulting from  wave-function factorization,  are worth being stressed:
\begin{enumerate}
\item  in the region of  relative distances
 $r_{ij} \gtrsim 1-1.5\,fm^{-1}$, nucleons "i" and "j"
move  independently in a mean field, with average relative momentum
 $k_{rel} \lesssim 1.5-2.0 \, fm^{-1}$, without any particular
difference between $pp$ and $pn$ distributions, apart from those due to the coulomb interaction; however,  as soon  as the relative
distance decreases down to a value of $r_{ij} \lesssim 1-1.5\,fm^{-1}$, the two nucleons   start feeling  the details of the NN
interaction, in particular  the tensor force which makes the $pn$ and $pp$ motions to appreciably differ, with the difference decreasing at
shorter distances, where  the strong NN repulsive part of the local NN interaction dominates. In the SRC regions, characterized by a large
content of high momentum components, thanks to the decoupling of the c.m. and the relative
 motions, also the two-nucleon momentum
distribution, independently of the mass of the nucleus, factorizes into a relative
and a c.m. parts; in particular,  in the
case of $pn$ pairs one has
 $n_A^{pn}(k_{rel}, K_{c.m.}, {\Theta})\simeq C_A^{pn} n_{D}^{{pn}}(k_{rel})
 {n_{c.m.}^{pn}}(K_{c.m.})$, where
$C_A^{pn}$ is an A-dependent constant, the {\it nuclear contact},
 which counts the number of deuteron-like  pairs  in nucleus $A$,  and
$n_D(k_{rel})$ is the deuteron momentum distribution;
we have shown that  the  deuteron-like factorized form  is valid only  at low values
of the c.m. momentum,
 $K_{c.m.} \lesssim 1-1.5 \, fm^{-1}$ and, at the same time, at high values
of the relative pair momentum
 $k_{rel} > k_{rel}^{-} \simeq 2 \, fm^{-1}$,  with the  value of  $k_{rel}^-$  increasing   with the value of
 $K_{cm}$; thus, the dynamics of in-medium $pn$ pairs  can, to a large extent,  be described as the dynamics of the
 motion in the nucleus of a deuteron-like pair, whose c.m. moves
 with a momentum distribution ${n_{c.m}^{pn}}(K_{c.m.})$;
\item within the above picture,
  arising from the factorization property of the momentum distributions,  the ratio
  $n_A^{{pn}}(k_{rel},K_{c.m.}=0)/
  [n_D(k_{rel}){n_{c.m.}^{pn}}(K_{c.m.}=0)]$ at high relative values
  of $k_{rel}$  should become a constant equal to $C_A^{pn}$, which is indeed the
  case; thus
  the theoretical values of the contacts $C_A^{pn}$, which have been determined by plotting the ratio {\it vs} $k_{rel}$,
  are completely free from any adjustable phenomenological parameter, for they are entirely defined in terms
  of   many-body quantities that are fixed by the choice of the NN interaction and by the
  way  the many-body problem is solved. This is true  for all nuclei considered,
  both within our cluster expansion approach and the VMC {\it ab initio} calculation.
  The  values
  of $C_A^{pn}$ range from about $2$ in $^3$He to about $60$ in $^{40}$Ca; for $^4$He the value of $C_A^{pn}$ is less
  by about 20\% than the value obtained
  with the VMC momentum distribution; such a difference should be ascribed  both to the different Hamiltonian (V8' NN interaction in our case and
  AV18 in Ref. \cite{Wiringa:2013ala}) and to the different variational wave functions;
  this point is under quantitative investigation;
\item
  for all nuclei that have been considered we found that  when
  $k_{rel}\gtrsim 2\,fm^{-1}$,
  the ratio $n_A^{{pn}}(k_{rel}, K_{c.m.}=0)/[C_A^{pn}{n_{c.m.}^{pn}}(K_{c.m.})]$
  practically does not differ from the deuteron momentum distribution
  $n_D(k_{rel})$, which is a further  clear evidence  of factorization
  of $n_A^{pn}(k_{rel}, K_{c.m.}, {\Theta})$; factorization also occurs
  when the numerator of the ratio  is replaced by the
  $K_{c.m.}$-integrated momentum distribution $n_A^{{pn}}(k_{rel})$, obtaining the ratio
  $n_A^{{pn}}(k_{rel})/[C_A^{pn}{n_{c.m.}^{pn}}(K_{c.m.})]$; this however is only true at very high values of
  $k_{rel}\gtrsim 4\, fm^{-1}$; this  means that at $k_{rel}\gtrsim 4\,fm^{-1}$
  $n_A^{{pn}}(k_{rel})$
  is dominated by the deuteron-like components with $K_{c.m.}=0$, whereas at lower values of $K_{c.m.}$ also the c.m. components with
  $K_{c.m.} \neq 0$ contribute;
\item we have considered the relationships between the one-nucleon and the
  two-nucleon momentum distribution, a topic
  recently discussed in Ref.
  \cite{Weiss:2015mba}. To this end we have compared three different approaches,
  namely: (i) the one in which only back-to-back ($K_{c.m.}=0$)
  correlated nucleons are considered; (ii) the convolution model developed in
  Ref. \cite{CiofidegliAtti:1995qe}; (iii)  the approach of Ref.
  \cite{Weiss:2015mba}, where the two-nucleon momentum distributions are considered
  in the asymptotic limit $k_1>>K_{c.m.}$; our results demonstrate
  that in all of  the three approaches the one-nucleon momentum distribution can be expressed, to a large extent,  in terms of a proper sum
  of the $pn$ and $pp$ distributions,  starting from a value of the one-nucleon momentum $k_1\gtrsim 2 \, fm^{-1}$, within approaches (i) and
  (ii) and starting at  $k_1\gtrsim 4 \, fm^{-1}$, within   approach (iii) ;
\item the two-nucleon momentum distributions have been used to calculate the absolute
  values of the number of SRCd  $pn$ and $pp$ pairs
  in the considered nuclei; in particular we have calculated the number of BB SRCd pairs, defined by the integral  of the two-nucleon
  momentum distribution in correspondence of $K_{c.m.}=0$ and (similar to the  deuteron case)
  in the relative momentum range $1.5
  <k_{rel}<\infty \,fm^{-1}$, finding in complex nuclei a number of {BB SRCd} $pn$ {\it pairs} larger than the number of
  $pp$ pairs by about  a factor of 10;
  concerning the numbers of SRCd  {\it nucleons} it should be stressed that
  that in our approach the  one- and two-nucleon momentum distribution   satisfy the exact
  relationship provided by Eq.(\ref{enne1_general}), which is  valid in the entire region of momentum
  $0 < k_{1}< \infty \,fm^{-1}$, so that the obtained two-nucleon momentum distributions
  provide a percent  ratio of  SRCd nucleons to the total number of nucleons in the range of
  16-20 \%,
  if SRCs are defined with respect to a pure independent-particle shell-model description.
\item The dependence upon $k_{rel}$ and $K_{c.m.}$ of the two-nucleon momentum distributions of ${^4}$He and $^{12}$C in the region of SRCs
  is in good agreement with available experimental data
  \cite{Tang:2002ww,Piasetzky:2006ai,Shneor:2007tu,Subedi:2008zz,Korover:2014dma,Hen:2014nza},  and so are  the c.m. distributions.
\end{enumerate}
Several aspects of the above picture, which we have shown to  occur also in
 light nuclei ($A \leq 12$) treated within the
VMC approach \cite{Wiringa:2013ala},  have already been experimentally confirmed, whereas some others, concerning in particular the values
of the nuclear contact in various spin-isospin states, deserve further theoretical and experimental investigations. Finally, we would like
to stress that  our approach provides  momentum distributions that in some momentum regions are  lower by 15-20 \% than the ones calculated
with the VMC momentum distributions; as already pointed out,  this can be attributed partly to the different Hamiltonian used in the two
approaches, and partly to the different variational wave functions; this point is under investigation.
  To conclude,  our approach
 turned out to be accurate enough to describe the main features of SRCs in few-nucleon systems and iso-scalar nuclei  with $A \leq 40$,
 so that
it should deserve the extension to different types of NN interaction models differing, particularly, in the short range behavior,  and
should be applied  to heavier  neutron-rich nuclei, whose investigation
presents several interesting aspects \cite{Hen:2014nza,Sargsian:2012sm}.\\
\vspace{1.5cm}
\section{Acknowledgments}\label{Sec:8}This work has been performed within the activity  and support of the Theory Group of the Italian
Istituto Nazionale di Fisica Nucleare (INFN), Sezione Perugia. H. M. is grateful to INFN for kind hospitality and support. We gratefully acknowledge
Robert Wiringa for providing useful information about the effects of the 3N force
 on the high momentum
components,
within the VMC approach.
M.A. acknowledges the CINECA award under the ISCRA initiative, for the availability of high
performance computing resources.

\end{document}